\documentclass[apj]{emulateapj}
\slugcomment{{\sc Accepted to ApJ:} March 18, 2011}
\usepackage{graphicx}
\usepackage[nointegrals]{wasysym}

\usepackage{verbatim}
\usepackage{amsmath}

\begin{document}

\title{Global Self-Similar Protostellar Disk/Wind Models}

\author{Seth Teitler\altaffilmark{1,2}}
\altaffiltext{1}{Presented as a dissertation to the Department of Astronomy and Astrophysics, The University of Chicago, in partial fulfillment of the requirements for the Ph.D. degree.}
\altaffiltext{2}{Current address: Department of Astronomy, University of Wisconsin--Madison, 475 N. Charter St., Madison, WI 53706, USA; email: steitler@astro.wisc.edu}

\affil{Department of Astronomy \& Astrophysics, University of Chicago, 5640 S Ellis Ave, Chicago, IL 60637, USA} 

\begin{abstract}
The magnetocentrifugal disk wind mechanism is the leading candidate for
producing the large-scale, bipolar jets commonly seen in protostellar
systems. I present a detailed formulation of a global, radially self-similar model for a non-ideal disk that launches a magnetocentrifugal wind.  This formulation generalizes the conductivity tensor formalism previously used in radially localized disk models.  The model involves matching a solution of the equations of non-ideal MHD describing matter in the disk to a solution of the equations of ideal MHD
describing a ``cold'' wind.  The disk solution must pass
smoothly through the sonic point, the wind solution must pass smoothly through the Alfv\'en point, and the two solutions must
match at the disk/wind interface.  This model includes for
the first time a self-consistent treatment of the evolution of magnetic
flux threading the disk, which can change on the disk accretion
timescale.  The formulation presented here also allows a realistic conductivity profile for the disk to be used in a global disk/wind model for the first time.  The physical constraints on the model solutions fix the distribution of the magnetic field
threading the disk, the midplane accretion speed, and the midplane
migration speed of flux surfaces.  I present a representative solution that corresponds to a disk in the ambipolar conductivity regime with a nominal neutral-matter--magnetic-field coupling parameter that is constant along field lines, matched to a wind solution.  I conclude with a brief discussion of the importance of self-similar disk/wind models in studying global processes such as dust evolution in protostellar systems.
\end{abstract}


\section{INTRODUCTION}\label{intro}

Low-mass star formation typically involves the collapse of a dense core
inside a molecular cloud to form a condensed central object
and an extended disk.  The protostar gains much of its mass by accretion
of disk material, which can only proceed if there is some mechanism for
removing angular momentum from the accreting material.  Protostars that
are actively accreting disk material frequently exhibit associated
bipolar outflows in the form of collimated jets oriented
perpendicular to the disk, stretching out over distances much larger
than the radial scale of the disk \citep[e.g.,][]{ba07}.  This
association between accretion disks and large-scale, collimated jets is
also seen in other systems, including X-ray binaries and active galactic
nuclei.  The strongest candidate mechanism for launching jets in these
systems is described by the magnetocentrifugal disk wind model
\citep[][hereafter BP82]{bp82}, which relies on the acceleration of a
fraction of disk material out along a large-scale, ordered magnetic
field threading the disk at a sufficiently small inclination angle.  This magnetic field could be the interstellar
field in the parent molecular cloud that is dragged in during core
collapse, or it could be generated by dynamo activity in the disk or the star.  A
natural consequence of the disk wind model is that the wind efficiently
removes angular momentum from the disk material, which could explain the
strong inferred link between accretion and outflow phenomena in protostars
\citep[e.g.,][]{ks11}.  

Several previous semianalytic studies have matched solutions of the equations of
non-ideal MHD that describe a weakly ionized, magnetized accretion disk,
and solutions of the equations of cold, ideal MHD for a polytropic fluid that
describe a BP82-type magnetocentrifugal wind that removes all the disk's
excess angular momentum.  These models focused on a radially
localized portion of the disk (e.g., \citealt{wk93}, hereafter WK93; \citealt{skw07}, hereafter SKW07; \citealt{skw11}, hereafter SKW11), or on a global disk model (e.g., \citealt{k89}; \citealt{li96}, hereafter L96;
\citealt{f97}, hereafter F97) based on the same assumption of
self-similarity in spherical radius that underlies the BP82 wind model.  Another approach is presented in \citet{ca03, ca05}, where the thinness of the disk is used to justify solutions for the disk structure that are separable in the radial and vertical coordinates; these solutions are matched to an isothermal wind.  A key parameter in determining whether a disk can launch a magnetocentrifugal wind is the field line inclination angle at the disk surface $B_{r,{\rm s}}/B_{z,{\rm s}}$, where $B_{r,{\rm s}}$ and $B_{z,{\rm s}}$ are the radial and vertical field components at the disk surface (subscript `s').  This inclination angle is determined by the flux distribution along the disk surface and can only be calculated in a self-consistent fashion in a global treatment \citep[][hereafter OL01]{ol01}.  A second key parameter describing the interaction of the disk and wind is $B_{\phi,{\rm s}}$, the azimuthal field strength at the disk surface, which sets the torque exerted by the wind on the disk (BP82).  As with the surface field inclination angle, the value of $B_{\phi,{\rm s}}$ is determined by processes acting outside the disk; specifically, the value of $B_{\phi,{\rm s}}$ is determined by the constraint that the wind solution pass smoothly through the Alfv\'en point.

Accretion flows in a weakly ionized disk will drag in magnetic flux, opposing the natural tendency of the magnetic flux to diffuse outwards.  Studies of accretion disks that shed angular momentum due to MHD turbulence developed from the magnetorotational instability (MRI, \citealt{bh91}) suggest that the magnetic diffusivity and the effective viscosity of such disks are comparable \citep{gg09, ll09}, agreeing with previous work by \citet{lpp94} which suggested that the magnetic diffusivity of MRI-active accretion disks may be too large to allow efficient inward dragging of magnetic flux.  This problem can be overcome in an MRI-active disk threaded by a large-scale field that removes angular momentum from surface layers in which the MRI is suppressed \citep{rl08, lrb09}.  More generally, this problem disappears if the angular momentum transport mechanism is decoupled from the mechanism that provides the magnetic diffusivity so that the diffusivity is much smaller than the effective viscosity of the disk.  This occurs naturally in a disk from which angular momentum is removed through a magnetocentrifugal wind, allowing magnetic flux to be dragged in and distributed over a wind-driving disk.  As noted by OL01, the evolution of the magnetic flux through the disk determines whether magnetic configurations capable of launching outflows can be maintained over the accretion timescale.  Previous studies of wind-driving disks have not treated the effects of magnetic flux evolution in a self-consistent manner: in a local model, OL01 calculated the rate of flux migration given $B_{r,{\rm s}}$ and $B_{\phi,{\rm s}}$, which were free parameters.  The local models of \citet{ca03, ca05} expressed $B_{\phi,{\rm s}}$ and the surface field inclination angle in terms of unconstrained parameters that describe the structure of the disk and the processes operating within it.  WK93 also employed a local model in which they determined $B_{\phi,{\rm s}}$ self-consistently using the Alfv\'en point constraint, but left the surface field inclination angle unconstrained, describing the magnetic flux migration speed using the free parameter $\epsilon_{\rm B}$.  The global models of L96 and F97 also solved for $B_{\phi,{\rm s}}$ using the Alfv\'en point constraint but left the surface field inclination angle and magnetic flux migration unconstrained, essentially setting $\epsilon_{\rm B} = 0$.  For an overview of recent work on semianalytic models and complementary numerical simulations of protostellar jets, see the article by \citet{ks11}.

I present here a new, semianalytic, radially self-similar, global
model that matches disk solutions with BP82-type wind solutions, adopting an approach similar to that of L96 but
including for the first time a fully self-consistent calculation of the
effects of magnetic flux migration and the magnetic field structure above the disk: $\epsilon_{\rm B}$, $B_{r,{\rm s}}$, $B_{\phi, {\rm s}}$ and  $B_{r,{\rm s}}/B_{z,{\rm s}}$ are no longer free parameters but are determined by physical
constraints imposed at the disk surface and at critical points of the
outflow.  In addition, the equations describing the disk are formulated
in terms of the conductivity tensor, allowing for fully general
non-ideal disk models.  The plan of the paper is as follows.  In
Section~\ref{form} I review the equations describing a (nearly) steady-state,
axisymmetric, weakly ionized disk threaded by a magnetic field, and I
show how the system reduces to a set
of ordinary differential equations under the radial self-similarity
assumption.  In Section~\ref{paramconstr} I discuss constraints from the self-consistent
treatment of the field-line inclination at the top of the disk and from
matching disk and wind solutions. In
Section~\ref{res} I present a representative matched disk/wind solution, and in
Section~\ref{concs} I summarize the study and outline plans to improve on the
current model.

\section{FORMULATION}\label{form}

I consider a weakly ionized, geometrically thin, vertically isothermal, axisymmetric disk in near-Keplerian rotation around a protostar of mass $M$ that is
threaded by an ordered, large-scale magnetic field.  I examine the general, non-ideal MHD case (Section~\ref{casegen}), expressing the matter--field interaction using the conductivity tensor formalism, and seek
solutions that are steady on timescales shorter than the accretion time
(Section~\ref{goveqgen}).  I follow L96 and F97 in making the
simplifying assumption of radial self-similarity to produce a global
disk/wind model (Section~\ref{selfgen}), and I derive the midplane values of various physical quantities (Section~\ref{midgen}).

In Section~\ref{caseAD} I specialize to the case of the ambipolar conductivity regime.  I once again derive the relevant self-similar disk equations (Section~\ref{selfAD}) and the midplane values for the variables of integration (Section~\ref{midAD}).

\subsection{General Non-Ideal Case}\label{casegen}

The interiors of protostellar disks have a low ionization fraction, necessitating the use of non-ideal MHD to describe the behavior of the disk material in the presence of a large-scale magnetic field that is assumed to thread the disk.  The non-ideal matter-field interaction can be described using a multi-fluid approach or by introducing the conductivity tensor \citep[e.g.,][]{ks11}, which encapsulates the response of the dominant neutral fluid to the presence of the magnetic field.  The effects of the magnetic field are communicated to the neutrals by the development of charge--neutral drift velocities due to the Lorentz forces on charged species, and the consequent systematic charge--neutral drag forces.  Under the conductivity tensor formalism, this is expressed in the generalized version of Ohm's law,
\begin{equation}
\mbox{\boldmath$J$} = \mbox{\boldmath$\sigma$} \cdot \mbox{\boldmath$E$}_{\rm c} = \sigma_{\rm O} \mbox{\boldmath$E$}_{\rm c\parallel} + \sigma_{\rm H} \hat {\mbox{\boldmath$B$}} \times \mbox{\boldmath$E$}_{\rm c\perp} + \sigma_{\rm P} \mbox{\boldmath$E$}_{\rm c\perp} \,,
\label{eq:ohm}
\end{equation}

\noindent where \boldmath$E$\unboldmath$_{\rm c}$ is the electric field in a frame comoving with the dominant neutral fluid, \boldmath$E$\unboldmath$_{\rm c\parallel}$ and \boldmath$E$\unboldmath$_{\rm c\perp}$ are the components of \boldmath$E$\unboldmath$_{\rm c}$ parallel and perpendicular to the magnetic field, respectively, and \boldmath$\sigma$\unboldmath is the conductivity tensor.  The components of \boldmath$\sigma$\unboldmath are the Ohm conductivity $\sigma_{\rm O}$, the Hall conductivity $\sigma_{\rm H}$, and the Pedersen conductivity $\sigma_{\rm P}$.

For a given charged species (denoted by subscript `$j$'), the product of the gyrofrequency $\left(\frac{|Z_{j}|e|B|}{m_{j}c}\right)$ and the neutral-charge momentum exchange time $1/\left(\gamma_{j}\rho\right)$ gives a measure of the coupling of the charged species to the magnetic field in a sea of neutrals.  In these expressions, $Z_{j}$, $m_{j}$, and $\gamma_{j}$ are the signed charge, mass, and charge--neutral collisional drag coefficient of species $j$, and $\rho$ is the neutral mass density.  The signed value of the ratio of the gyrofrequency and neutral collision frequency is the Hall parameter,

\begin{equation}
\beta_{j} \equiv \frac{|Z_{j}|eB}{m_{j}c} \frac{1}{\gamma_{j}\rho}\,.
\label{eq:hallparam}
\end{equation}

\noindent The definition of the Hall parameter uses the signed value of the magnetic field strength, $B \equiv |\mbox{\boldmath$B$}|sgn\{B_{z}\}$, where $B_{z}$ is the $z$ component of the magnetic field in cylindrical coordinates $\{r, \phi, z\}$, so that the Hall conductivity retains its dependence on the magnetic field polarity (see Equation (\ref{eq:sigmahall})).  Large values $|\beta|_{j} \gg 1$ imply that charged species $j$ is tightly coupled to the magnetic field, whereas small values $|\beta_{j}| \ll 1$ imply that species $j$ is weakly coupled to the field.  

The conductivity tensor components can be expressed in terms of the Hall parameters as

\begin{equation}
\sigma_{\rm O} = \frac{ec}{B}\sum\limits_{j}n_{j}|Z_{j}|\beta_{j}\,,
\label{eq:sigmaohm}
\end{equation}

\begin{equation}
\sigma_{\rm H} = \frac{ec}{B}\sum\limits_{j}\frac{n_{j}|Z_{j}|}{1+\beta_{j}^{2}} \,,
\label{eq:sigmahall}
\end{equation}

\noindent and

\begin{equation}
\sigma_{\rm P} = \frac{ec}{B}\sum\limits_{j}\frac{n_{j}|Z_{j}|\beta_{j}}{1+\beta_{j}^{2}} \,.
\label{eq:sigmapedersen}
\end{equation}

\noindent Note that since $B$ is the signed value of the magnetic field strength, the Hall conductivity can be positive or negative.  The factor of $B$ in $\beta_{j}$ cancels the pre-factor of $1/B$ in the expressions for the Ohm and Pedersen conductivities, which are positive regardless of the magnetic field orientation.

The coupling of the neutral matter to the magnetic field is described by the Elsasser number $\Lambda$, given by
\begin{equation}
\Lambda \equiv \frac{v_{\rm A}^{2}}{\Omega_{\rm K}\eta_{\perp}} \,,
\label{eq:elsasser}
\end{equation}

\noindent where $v_{\rm A} \equiv |\mbox{\boldmath$B$}|/\sqrt{4 \pi \rho}$ is the Alfv\'en speed, $\Omega_{\rm K}$ is the Keplerian angular frequency, $\eta_{\perp} = c^{2}/4\pi \sigma_{\perp}$, and $\sigma_{\perp} = (\sigma_{\rm H}^{2}+\sigma_{\rm P}^{2})^{1/2}$ \citep[e.g.,][]{ks11}.  As with the Hall parameter, large values $\Lambda \gg 1$ imply that the neutral matter is strongly coupled to the magnetic field, whereas small values $\Lambda \ll 1$ imply that the neutral matter is only weakly coupled to the magnetic field.  Launching a magnetocentrifugal wind typically requires strong neutral--field coupling ($\Lambda \gtrsim 1$).

\subsubsection{Governing Equations}\label{goveqgen}

The governing equations for the disk are the continuity of mass, 
\begin{equation} 
\nabla \cdot \left(\rho \mbox{\boldmath$V$}\right) = 0 \,,
\label{eq:mass} 
\end{equation}
and the conservation of momentum, 
\begin{equation}
\frac{\partial \mbox{\boldmath$V$}}{\partial t} + \left(\mbox{\boldmath$V$} \cdot
\nabla\right) \mbox{\boldmath$V$} + \frac{1} {\rho} \nabla P + \nabla
\Phi - \frac{\mbox{\boldmath$J$} \times \mbox{\boldmath$B$}}{c \rho} = 0 \,.
\label{eq:moment} 
\end{equation}
In these equations, \boldmath$V$\unboldmath is the neutral gas velocity and $P = \rho c_{\rm
s}^2$ is its pressure, where $c_{\rm s}$ is the isothermal speed of
sound. The gravitational potential of the central object is given by
$\Phi = -GM/R = -V_{\rm K}^2(R)$, where $R = (r^2 + z^2)^{1/2}$ is the
spherical radius and $V_{\rm
K}$ is the Keplerian rotation speed.

Within the disk, the ionization fraction is very low, which implies that the equations of momentum conservation for the charged species are dominated by the charge--neutral drag force and the Lorentz force (cf. Equations 2.15--2.17 in WK93).  We can thus replace the charge--neutral drag force with the Lorentz force; this substitution has already been made in the conservation of momentum for the neutral disk material (Equation (\ref{eq:moment})).

The current density satisfies Amp\`{e}re's law (neglecting the
displacement current)
\begin{equation}
\mbox{\boldmath$J$} = \frac{c}{4\pi} \nabla \times \mbox{\boldmath$B$} \,,
\label{eq:amp}
\end{equation}

\noindent and Ohm's law (Equation (\ref{eq:ohm})), with the comoving electric field \boldmath$E$\unboldmath$_{\rm c}$ related to the inertial-frame electric field \boldmath$E$\unboldmath by 
\begin{equation}
\mbox{\boldmath$E$}_{\rm c} = \mbox{\boldmath$E$} + \frac{\mbox{\boldmath$V$} \times \mbox{\boldmath$B$}}{c} \,.
\label{eq:ec}
\end{equation}

The magnetic field satisfies the solenoidal condition
\begin{equation}
\nabla \cdot \mbox{\boldmath$B$} = 0 \,,
\label{eq:solen}
\end{equation}
and the electric field \boldmath$E$\unboldmath satisfies the induction equation, 
\begin{equation} 
\frac{\partial \mbox{\boldmath$B$}}{\partial t} = -c \nabla \times \mbox{\boldmath$E$} \,, 
\label{eq:induct} 
\end{equation}

\noindent or, in terms of the comoving electric field,
\begin{equation} 
\frac{\partial \mbox{\boldmath$B$}}{\partial t} = \nabla \times \left(\mbox{\boldmath$V$}
\times \mbox{\boldmath$B$}\right) -c \nabla \times \mbox{\boldmath$E$}_{\rm c} \,.
\label{eq:induct3} 
\end{equation}
In a strictly steady state, $\partial \mbox{\boldmath$B$}/\partial t = 0$
identically.  However, the magnetic flux threading the disk interior to
a fixed radius $r$ can change on the accretion timescale due to
advection of field lines by the accretion flow and diffusion of field
lines through the non-ideal disk material.  I therefore consider
disk solutions that are steady on timescales longer than the dynamic
time $\sim r/V_{\rm K}$ but relax the steady-state assumption on
timescales comparable to the accretion time $\sim r/|V_r|$.

Expanding Equation (\ref{eq:induct3}) yields 
\begin{align}
\frac{\partial B_{r}}{\partial t} = & V_{r} \frac{\partial B_{z}}{\partial
z} + B_{z} \frac{\partial V_{r}}{\partial z} - V_{z} \frac{\partial
B_{r}}{\partial z} \nonumber \\
& - B_{r} \frac{\partial V_{z}}{\partial z} + c
\frac{\partial E_{\rm c\phi}}{\partial z} \,,
\label{eq:inductr}
\end{align} 
\begin{align} 
\frac{\partial B_{\phi}}{\partial t} = & B_{r} \frac{\partial
V_{\phi}}{\partial r} + B_{z} \frac{\partial V_{\phi}}{\partial z} -
\frac{V_{\phi} B_{r}}{r} - V_{z} \frac{\partial B_{\phi}}{\partial z} - B_{\phi} \frac{\partial V_{z}}{\partial z} \nonumber \\
& - B_{\phi} \frac{\partial
V_{r}}{\partial r} - V_{r} \frac{\partial B_{\phi}}{\partial r} + c
\frac{\partial E_{{\rm c} z}}{\partial r} - c \frac{\partial E_{{\rm c} r}}{\partial
z}\,,
\label{eq:inductphi}
\end{align} 
and 
\begin{align} 
\frac{\partial B_{z}}{\partial t} = & \frac{V_{z} B_{r}}{r} - \frac{V_{r}
B_{z}}{r} + V_{z} \frac{\partial B_{r}}{\partial r} + B_{r}
\frac{\partial V_{z}}{\partial r} \nonumber \\
& - V_{r} \frac{\partial B_{z}}{\partial
r} - B_{z}\frac{\partial V_{r}}{\partial r} - c \frac{E_{\rm c\phi}}{r} - c
\frac{\partial E_{\rm c\phi}}{\partial r} \,,
\label{eq:inductz} 
\end{align} 

\noindent where the solenoidal condition (Equation (\ref{eq:solen})) was used to obtain
Equation (\ref{eq:inductphi}). Now, $V_{\phi} \approx V_{\rm K}$,
$\left|V_{r}\right| \approx c_{\rm s}$ through much of the disk and
$\left|V_{r}\right| \approx V_{z}$ near the top of the disk (see Figure 2), $z/r
\lesssim h_{\rm T}/r= c_{\rm s}/V_{\rm K} \ll 1$ (where $h_{\rm T}$ is
the tidal scale height), and $c\mbox{\boldmath$E$}_{\rm c} \approx \mbox{\boldmath$V$} \times \mbox{\boldmath$B$}$. These relations imply that the
terms $V_{r}~\partial B_{z}/\partial z$, $B_{z}~\partial V_{r}/\partial
z$, $V_{z}~\partial B_{r}/\partial z$, $B_{r}~\partial V_{z}/\partial
z$, and $c~\partial E_{\rm c\phi}/ \partial z$ in Equation (\ref{eq:inductr}) describe processes operating on the dynamic timescale.  Similarly, the terms $B_{r}~\partial
V_{\phi}/\partial r$, $B_{z}~\partial V_{\phi}/\partial z$, $V_{\phi}~B_{r}/{r}$, $V_{z}~\partial B_{\phi}/\partial z$, $B_{\phi}~\partial
V_{z}/\partial z$, $c~\partial E_{{\rm c}z}/\partial r$, and $c~\partial E_{{\rm c}r}/ \partial z$ in Equation (\ref{eq:inductphi}) are associated with
processes operating on the dynamic timescale.  However, every term in
Equation (\ref{eq:inductz}) describes a process that operates on the
longer accretion timescale.  I therefore ignore $\partial B_{r}/\partial
t$ and $\partial B_{\phi}/\partial t$ in the
$r$ and $\phi$ components of the induction equation, since those terms
represent changes in the magnetic field on the accretion timescale and
are negligible compared with the terms representing changes in the
magnetic field on the dynamic timescale, but I retain $\partial
B_{z}/\partial t$ in the $z$ component of the induction equation
\citep[cf.][]{ksw10}. 

The components of Equation (\ref{eq:induct}) thus become
\begin{equation} 
c\frac{\partial E_{\phi}}{\partial z} = 0 \,,
\label{eq:inductr2} 
\end{equation}

\begin{equation} 
c\frac{\partial E_{r}}{\partial z}-c \frac{\partial E_{z}}{\partial r} = 0 \,,
\label{eq:inductphi2} 
\end{equation}

\noindent and
\begin{equation} 
\frac{\partial B_{z}}{\partial t} = \frac{c}{r}\frac{\partial \left(rE_{\phi}\right)}{\partial r}\,.
\label{eq:inductz2} 
\end{equation}
The $r$ and $z$ components of the induction equation together imply that $E_{\phi}$ is constant with
height at a given $r$ but need not vanish identically.

\subsubsection{Self-Similarity Equations}\label{selfgen}

The disk equations reduce from partial to ordinary differential
equations under the assumption of radial self-similarity, which implies
that all quantities are described by power laws in the spherical radius $R$
for a fixed inclination angle.  I normalize all quantities by midplane
values on a reference magnetic field line (subscript `*'), using a scheme similar to that found in
\citet{li95} and L96.  Magnetic field lines can be labeled by the magnetic flux function $\Psi$, which satisfies $\nabla\Psi \times \hat{\phi} = B_{\rm p} r$, where $B_{\rm p}$ is the poloidal magnetic field.  Note that the notation used here corresponds to that of \citet{li95} and L96 but differs from that of \citet{ksw10}, where the flux function is labeled $A$ and $\Psi$ is used to denote the poloidal flux; the two quantities are related by $A = \Psi/2\pi$.

Under the self-similarity assumption, the midplane values of all
physical quantities scale as power laws in $\Psi$, and 
\begin{equation}
\Psi \propto r^{1/\zeta}\,.
\label{eq:psidef}
\end{equation}
\noindent The magnetic flux distribution parameter $\zeta$ is
denoted by $\xi$ in L96.  The cold-wind model of BP82 has $\zeta = 4/3$.
In a disk that drives a magnetocentrifugal wind, $\zeta < 4/3$, and if
the outflow rate in the wind is much smaller than the accretion rate, as
is the case in protostellar disks, then $\delta\zeta \equiv 4/3 - \zeta
\ll \zeta$ (L96).

In terms of the self-similarity dimensionless variables $\varphi \equiv \Psi/\Psi_{*}$ and $s \equiv z/r$ (denoted by $t$ in L96), various physical quantities are normalized as follows:

\begin{align}
& r = r_{*} \varphi^{\zeta} ~x\left(s \right); & z = r_{*} \varphi^{\zeta} ~s x\left(s \right); \nonumber \\
&& & \nonumber \\
& \rho = \rho_{*} \varphi^{2-3\zeta} ~\tilde{\rho}\left(s \right); & \mbox{\boldmath$V$} = V_{\rm K*} \varphi^{-\frac{\zeta}{2}} ~\mbox{\boldmath$v$}\left(s \right); \nonumber \\
&& & \nonumber \\
& \mbox{\boldmath$B$} = B_{z*} \varphi^{1-2\zeta} ~\mbox{\boldmath$b$}\left(s \right); & \mbox{\boldmath$E$} = \frac{V_{\rm K*}B_{z*}}{c}\varphi^{1-\frac{5}{2}\zeta} ~\mbox{\boldmath$e$}\left(s \right); \nonumber \\
& & & \nonumber  \\
& \mbox{\boldmath$E$}_{\rm c} = \frac{V_{\rm K*}B_{z*}}{c}\varphi^{1-\frac{5}{2}\zeta} ~\mbox{\boldmath$e$}_{\rm c}\left(s \right); & P = P_{*} \varphi^{2-4\zeta} ~p\left(s \right); \nonumber \\
&& & \nonumber \\
& \mbox{\boldmath$\sigma$} = \frac{c^{2}}{4\pi r_{*} V_{\rm K*}}\varphi^{-\frac{\zeta}{2}} ~\tilde{\mbox{\boldmath$\sigma$}}\left(s \right);  & \mbox{\boldmath$J$} =  \frac{cB_{z*}}{4\pi r_{*}}\varphi^{1-3\zeta} ~\mbox{\boldmath$j$}\left(s \right) \,.
\label{eq:nondim}
\end{align}

The variable $\varphi$ raised to various powers converts midplane values
 on the reference field line (subscript `*'), which is labeled by flux
 function value $\Psi_{*}$, to midplane values on a generic field line
 (subscript `0') with flux function value $\Psi = \Psi_{0}$.  The
 combination $r_{*} \varphi^{\zeta}$, for example, is equal to $r_{0}$,
 the midplane cylindrical radius of the field line labeled by flux
 function value $\Psi_{0}$.  The exponents for the power laws in
 $\varphi$ in Equation (\ref{eq:nondim}) come from dimensional analysis of the governing equations.  The variable $s$ expresses position along a field line.  The new coordinates $\{s\,,\,\varphi\}$ are expressed in terms of the spatial coordinates $\{r\,,\,z\}$ as $s=z/r$ and $\varphi=\left(\frac{1}{r_{*}}\frac{r}{x\left(z/r\right)}\right)^{1/\zeta}$.  This gives

\begin{align}
\left(\frac{\partial s}{\partial r}\right)_{z} = -\frac{s}{r}; & \left(\frac{\partial s}{\partial z}\right)_{r} =  \frac{1}{r};  \nonumber \\
&  \nonumber \\
\left(\frac{\partial \varphi}{\partial r}\right)_{z} =  \frac{1}{\zeta}\frac{\varphi}{r}\left(\frac{x+x^{\prime}s}{x}\right); & \left(\frac{\partial \varphi}{\partial z}\right)_{r} = \frac{1}{\zeta}\frac{\varphi}{r}\left(-\frac{x^{\prime}}{x}\right) \,.
\end{align}

\noindent In these relations, $x^{\prime} \equiv \left(\frac{\partial x}{\partial s}\right)_{\varphi}$.  In terms of the new coordinates, the spatial derivatives of a generic quantity $A = A_{*} \varphi^{\alpha_{a}} a\left(s\right)$ take the following forms \citep{li95}:

\begin{align} 
\left(\frac{\partial A}{\partial r}\right)_{z} & = \left(\frac{\partial A}{\partial s}\right)_{\varphi} \left(\frac{\partial s}{\partial r}\right)_{z} + \left(\frac{\partial A}{\partial \varphi}\right)_{s} \left(\frac{\partial \varphi}{\partial r}\right)_{z}  \nonumber \\
\nonumber \\
& = \frac{A_{*}\varphi^{\alpha_{a}}}{r_{*}\varphi^{\zeta}}\frac{\alpha_{a}a\left(x+x^{\prime}s\right)-\zeta a^{\prime}xs}{\zeta x^{2}}
\label{eq:derr} 
\end{align}

\noindent and

\begin{align}
\left(\frac{\partial A}{\partial z}\right)_{r} & = \left(\frac{\partial A}{\partial s}\right)_{\varphi} \left(\frac{\partial s}{\partial z}\right)_{r} + \left(\frac{\partial A}{\partial \varphi}\right)_{s} \left(\frac{\partial \varphi}{\partial z}\right)_{r}  \nonumber \\
& = \frac{A_{*}\varphi^{\alpha_{a}}}{r_{*}\varphi^{\zeta}}\frac{\zeta a^{\prime}x-\alpha_{a}ax^{\prime}}{\zeta x^{2}} \,,
\nonumber \\
\label{eq:derz}
\end{align}

\noindent where $a^{\prime} \equiv \left(\frac{\partial a}{\partial s}\right)_{\varphi}$.

The partial differential equations describing the disk can now be recast as ordinary differential equations and algebraic relations in the normalized quantities.  The equation of mass continuity becomes

\begin{align} 
v_{z}^{\prime}-sv_{r}^{\prime}+\frac{v_{z}-v_{r}s }{\tilde{\rho}}\tilde{\rho}^{\prime} & = \left(\frac{5}{2}-\frac{2}{\zeta}\right)v_{r} \nonumber \\
& + \left(\frac{2}{\zeta}-\frac{7}{2}\right)\frac{x^{\prime}}{x}(v_{z}-v_{r}s )\,,
\label{eq:mass2} 
\end{align}

\noindent and the conservation of momentum (Equation (\ref{eq:moment})) becomes
\begin{align}
\left(v_{z}-v_{r}s \right)&v_{r}^{\prime}-\frac{\theta s}{\tilde{\rho}}\tilde{\rho}^{\prime} = \frac{v_{r}^{2}}{2}+v_{\phi}^{2}-\left(v_{z}-v_{r}s \right)\frac{v_{r}x^{\prime}}{2x} \nonumber \\
& -\frac{1}{\left(1+s^{2}\right)^{3/2} }+\left(4-\frac{2}{\zeta}\right)\theta\left(1+\frac{x^{\prime}s }{x}\right) \nonumber \\
&+\frac{\nu x}{\tilde{\rho}}\left(j_{\phi}b_{z}-j_{z}b_{\phi}\right) \label{eq:mom2r}
\end{align}

\noindent in the $r$ direction,

\begin{align}
\left(v_{z}-v_{r}s \right)v_{\phi}^{\prime}=& -\frac{v_{r}v_{\phi}}{2}-\left(v_{z}-v_{r}s \right)\frac{v_{\phi}x^{\prime}}{2x} \nonumber \\
&+\frac{\nu x}{\tilde{\rho}}\left(j_{z}b_{r}-j_{r}b_{z}\right)
\label{eq:mom2phi}
\end{align}

\noindent in the $\phi$ direction, and

\begin{align}
&\left(v_{z}-v_{r}s \right)v_{z}^{\prime}+\frac{\theta}{\tilde{\rho}}\tilde{\rho}^{\prime} = \frac{v_{r}v_{z}}{2}-\left(v_{z}-v_{r}s \right)\frac{v_{z}x^{\prime}}{2x} \nonumber \\
&-\frac{s}{\left(1+s^{2}\right)^{3/2} }-\left(4-\frac{2}{\zeta}\right)\theta\frac{x^{\prime}}{x}+\frac{\nu x}{\tilde{\rho}}\left(j_{r}b_{\phi}-j_{\phi}b_{r}\right) \label{eq:mom2z}
\end{align}

\noindent in the $z$ direction, where 

\begin{equation}
\theta \equiv c_{\rm s}^{2}/v_{\rm K}^{2}
\label{eq:thetadef}
\end{equation}

\noindent and

\begin{equation}
\nu \equiv v_{\rm A,0}^{2}/v_{\rm K,0}^{2}\,.
\label{eq:nudef}
\end{equation}
The quantity $v_{\rm A,0}^{2}/v_{\rm K}^{2}$ is denoted by $\sigma$ in L96.

By definition, $\Psi = \Psi_{*}\varphi$.  The relation $B_{r}r =
-\left(\frac{\partial \Psi}{\partial z}\right)_{r}$ therefore gives
$\left(B_{z*} \varphi^{1-2\zeta} b_{r}\right)
\left(r_{*}\varphi^{\zeta}x\right) =
\frac{\Psi_{*}\varphi}{r_{*}\varphi^{\zeta}}\left(\frac{x^{\prime}}{\zeta
x^{2}}\right)$.  From the definition of the flux function,
$\frac{\Psi_{*}}{r_{*}^{2}} = B_{z*}$, so we have

\begin{equation}
x^{\prime}  = b_{r}x^{3} \,.
\label{eq:xprime}
\end{equation}

\noindent At a point $\left(r,z\right) = r_{*}\varphi^{\zeta}\left(x(s),sx(s)\right)$ on a field line, the unit tangent vector along the field line is 

\begin{displaymath}
\frac{x^{\prime}\hat{r}+\left(x+x^{\prime}s \right)\hat{z}}{\sqrt{x^{\prime 2} + \left(x+x^{\prime}s \right)^{2}}} = \frac{b_{r}\hat{r}+b_{z}\hat{z}}{\sqrt{b_{r}^{2}+b_{z}^{2}}} \,.
\end{displaymath}

\noindent This relation, combined with Equation (\ref{eq:xprime}), gives 
\begin{equation}
b_{z} = b_{r}s+\frac{1}{x^{2}} \,.
\label{eq:brbz}
\end{equation}
\noindent Equation (\ref{eq:brbz}), which expresses $b_{z}$ in terms of $b_{r}$, $x$, and $s$, automatically satisfies the solenoidal condition (Equation (\ref{eq:solen})).

Amp\`{e}re's law (Equation (\ref{eq:amp})) becomes

\begin{equation}
b_{\phi}^{\prime} = \left(\frac{1}{\zeta}-2\right)\frac{x^{\prime}}{x}b_{\phi} - x j_{r}
\label{eq:ampr2}
\end{equation}

\noindent in the $r$ direction,

\begin{equation}
b_{r}^{\prime} + sb_{z}^{\prime} = x j_{\phi} + \left(\frac{1}{\zeta}-2\right)\left[\frac{x^{\prime}}{x}\left(b_{r}+b_{z}s\right)+b_{z}\right]
\label{eq:ampphi3}
\end{equation}

\noindent in the $\phi$ direction, and

\begin{equation}
sb_{\phi}^{\prime} =  b_{\phi}+\left(\frac{1}{\zeta}-2\right)b_{\phi}\left(1+\frac{x^{\prime}s }{x}\right)-x j_{z}
\label{eq:ampz2}
\end{equation}

\noindent in the $z$ direction.  Using the algebraic relation for $b_{z}$, the $\phi$ component simplifies further to

\begin{equation}
b_{r}^{\prime} = \frac{x j_{\phi} + b_{r}s + \left(\frac{1}{\zeta}-2\right)\left[\frac{x^{\prime}}{x}\left(b_{r}+b_{z}s\right)+b_{z}\right]}{1+s^{2}} \,.
\label{eq:ampphi2}
\end{equation}

\noindent The poloidal components of Amp\`{e}re's law (Equations (\ref{eq:ampr2}) and (\ref{eq:ampz2})) together imply

\begin{equation}
j_{z} = j_{r}s + \left(\frac{1}{\zeta}-1\right)\frac{b_{\phi}}{x}\,.
\label{eq:jrjz}
\end{equation}

Ohm's law  (Equation (\ref{eq:ohm})) becomes

\begin{equation}
\mbox{\boldmath$j$} = \tilde{\mbox{\boldmath$\sigma$}} \cdot \mbox{\boldmath$e$}_{\rm c} =  \tilde{\sigma}_{\rm O} \mbox{\boldmath$e$}_{\rm c\parallel} + \tilde{\sigma}_{\rm H} \hat{b} \times \mbox{\boldmath$e$}_{\rm c\perp} + \tilde{\sigma}_{\rm P} \mbox{\boldmath$e$}_{\rm c\perp} \,,
\label{eq:ohm2}
\end{equation}

\noindent where the comoving electric field is given by

\begin{equation}
\mbox{\boldmath$e$}_{\rm c} = \mbox{\boldmath$e$} + \mbox{\boldmath$v$} \times \mbox{\boldmath$b$} \,
\label{eq:ec2}
\end{equation}

\noindent The three components of Equation (\ref{eq:ohm2}) are

\begin{equation}
j_{r} = \tilde{\sigma}_{\rm O} e_{\rm{c}\it{r}\parallel} + \frac{\tilde{\sigma}_{\rm H}}{b}\left(b_{\phi}e_{\rm{c}\it{z}\perp}-b_{z}e_{\rm{c}\it{\phi}\perp}\right) + \tilde{\sigma}_{\rm P} e_{\rm{c}\it{r}\perp} \,,
\label{eq:ohm2r}
\end{equation}

\begin{equation}
j_{\phi} = \tilde{\sigma}_{\rm O} e_{\rm{c}\it{\phi}\parallel} + \frac{\tilde{\sigma}_{\rm H}}{b}\left(b_{z}e_{\rm{c}\it{r}\perp}-b_{r}e_{\rm{c}\it{z}\perp}\right) + \tilde{\sigma}_{\rm P} e_{\rm{c}\it{\phi}\perp} \,,
\label{eq:ohm2phi}
\end{equation}

\begin{equation}
j_{z} = \tilde{\sigma}_{\rm O} e_{\rm{c}\it{z}\parallel} + \frac{\tilde{\sigma}_{\rm H}}{b}\left(b_{r}e_{\rm{c}\it{\phi}\perp}-b_{\phi}e_{\rm{c}\it{r}\perp}\right) + \tilde{\sigma}_{\rm P} e_{\rm{c}\it{z}\perp} \,,
\label{eq:ohm2z}
\end{equation}

\noindent where $b = \sqrt{b_{r}^{2}+b_{\phi}^{2}+b_{z}^{2}}\,$, and the magnetic field-parallel and -perpendicular components of the comoving electric field satisfy

\begin{equation}
\frac{e_{\rm{c}\it{r}\parallel}}{b_{r}} = \frac{e_{\rm{c}\it{\phi}\parallel}}{b_{\phi}} = \frac{e_{\rm{c}\it{z}\parallel}}{b_{z}} \,,
\label{eq:ecpar}
\end{equation}

\begin{equation}
e_{\rm{c}\it{r}\perp}b_{r} + e_{\rm{c}\it{\phi}\perp}b_{\phi} + e_{\rm{c}\it{z}\perp}b_{z} = 0 \,,
\label{eq:ecperp}
\end{equation}

\noindent and

\begin{equation}
\mbox{\boldmath$e$}_{\rm c} = \mbox{\boldmath$e$}_{\rm c\parallel} + \mbox{\boldmath$e$}_{\rm c\perp} \,.
\label{eq:ecparperp}
\end{equation}

The poloidal components of the induction equation (Equations (\ref{eq:inductr2}) and (\ref{eq:inductz2})) yield the relation $e_{\phi}\left(r,z\right) = e_{\phi}\left(r,0\right)$, which implies

\begin{equation}
e_{\phi}(s) = e_{\phi}(0)x(s)^{\frac{1}{\xi} - \frac{5}{2}} \,.
\label{eq:induct2b}
\end{equation}

\noindent The $\phi$ component of the induction equation (Equation (\ref{eq:inductphi2})) yields
\begin{equation}
e_{r}^{\prime} + se_{z}^{\prime}=\left(\frac{1}{\xi}-\frac{5}{2}\right)\left[\frac{x^{\prime}}{x}\left(e_{r}+e_{z}s\right)+e_{z}\right] \,.
\label{eq:induct2a}
\end{equation}

Combining equations  (\ref{eq:mass2}), (\ref{eq:mom2r}), and (\ref{eq:mom2z}) yields

\begin{align}
\tilde{\rho}^{\prime} = &\left[\frac{\tilde{\rho}}{\left(v_{z}-v_{r}s  \right)^{2}-\theta \left(1+s^{2}\right)}\right] \left[\left(2-\frac{2}{\zeta}\right)v_{r}\left(v_{z}-v_{r}s \right)\right. \nonumber \\
& \left.+\left(\frac{2}{\zeta}-3\right)\frac{x^{\prime}}{x}\left(v_{z}-v_{r}s \right)^{2} \right. \nonumber \\
& \left. +\left(4-\frac{2}{\zeta}\right)\theta\left(s+\frac{x^{\prime}}{x}+\frac{x^{\prime}s ^{2}}{x}\right)+v_{\phi}^{2}s\right. \nonumber \\
& \left. +\frac{\nu x}{\tilde{\rho}}\left(j_{\phi}\left[b_{r}+b_{z}s\right]-b_{\phi}\left[j_{r}+j_{z}s\right]\right)\right]\,.
\label{eq:rhoprime}
\end{align}

\noindent This relation shows that the point where $\left(v_{z}-v_{r}s  \right)^{2} = \theta \left(1+s^{2}\right)$ is a singular point of the system of equations, since the denominator of $\tilde{\rho}^{\prime}$ vanishes there.  The quantity $\sqrt{\left(v_{z}-v_{r}s  \right)^{2}/\left(1+s^{2}\right)}$ is the component of the poloidal velocity perpendicular to the local surface of self-similarity (the surface of constant $s$), normalized by the Keplerian velocity.  The denominator of $\tilde{\rho}^{\prime}$ vanishes when $\left(v_{z}-v_{r}s  \right)^{2}/\left(1+s^{2}\right) = \theta = \left(c_{\rm s}/V_{\rm K}\right)^{2}$.  This occurs when the unnormalized velocity component $V_{z}-V_{r}s/\sqrt{1+s^{2}}$ becomes equal to the local sound speed; this defines the sonic point.  The Alfv\'en point is similarly a critical point of the cold ideal wind equations, and occurs where the same poloidal velocity component $V_{z}-V_{r}s/\sqrt{1+s^{2}}$ reaches the local Alfv\'en speed.

Combining Equation (\ref{eq:rhoprime}) with Equations (\ref{eq:mom2r}) and (\ref{eq:mom2z}) gives expressions for $v_{r}^{\prime}$ and $v_{z}^{\prime}$.  Equation (\ref{eq:xprime}) gives $x^{\prime}$, Equation (\ref{eq:mom2phi}) gives $v_{\phi}^{\prime}$, and Equations (\ref{eq:ampphi2}) and (\ref{eq:ampr2}) give $b_{r}^{\prime}$ and $b_{\phi}^{\prime}$, respectively.  The final ordinary differential equation is Equation (\ref{eq:induct2a}).  Solving these 8 ordinary differential equations requires knowledge of the conductivity profile $\tilde{\sigma}_{\rm O}(s)$, $\tilde{\sigma}_{\rm H}(s)$, and $\tilde{\sigma}_{\rm P}(s)$, which can be obtained using a known disk ionoization profile or some simplifying assumptions.  In order to solve for $e_{r}^{\prime}$ in Equation (\ref{eq:induct2a}), it is necessary to express $e_{z}$ as a function of the other variables and then differentiate the resulting relation.  This expression is derived in Appendix \ref{appez}.

\subsubsection{Values at the Midplane}\label{midgen}

All physical quantities are assumed to have either odd or even symmetry
in $z$ about the midplane. The magnetic field is vertical at the
midplane, so the $s$ derivative coincides with a $z$ derivative at that
location.

By definition, $x_{0} = 1$ and $\tilde{\rho}_{0} = 1$.  Since the
magnetic field is vertical at the midplane, $b_{r,0} = 0$ and
$b_{\phi,0} = 0$, and since the vertical component of the flow velocity
must be continuous across the midplane, $v_{z,0} = 0$.  The midplane
radial velocity and azimuthal electric field are given 
by 

\begin{equation}
\epsilon \equiv -V_{r,0}/c_{\rm s}
\label{eq:epsdef}
\end{equation}

\noindent and 

\begin{equation}
\epsilon_{\rm B} \equiv -cE_{\phi,0}/c_{\rm s}B_{z,0}\,.
\label{eq:epsbdef}
\end{equation}

In dimensionless form, these relations are $v_{r,0} = -\epsilon \theta^{1/2}$ and $e_{\phi,0} = -\epsilon_{\rm B} \theta^{1/2}$, which combines with Equation (\ref{eq:induct2b}) to
give

\begin{equation}
e_{\phi}(s) = -\epsilon_{\rm B} \theta^{1/2}x(s)^{\frac{1}{\xi} - \frac{5}{2}} \,.
\label{eq:induct3b}
\end{equation}

\noindent The quantities $\epsilon$ and $\epsilon_{\rm B}$ are
parameters of the model whose values are set by conditions that must be
satisfied at special points in the flow (see Section~\ref{paramconstr}).  The
definitions of $\epsilon$ and $\epsilon_{\rm B}$ given here are the same
as in WK93; a factor of $\theta^{1/2}$ appears in the expressions for
$v_{r,0}$ and $e_{\phi,0}$ because velocities here are normalized by
$v_{\rm K}$ rather than by $c_{\rm s}$ as in WK93.

At the midplane, Equations (\ref{eq:mom2r}) and (\ref{eq:mom2phi}) become
\begin{equation}
\frac{v_{r,0}^{2}}{2}+v_{\phi,0}^{2}-1+\left(4-\frac{2}{\zeta}\right)\theta+\nu j_{\phi,0} = 0 \,
\label{eq:mom2r0}
\end{equation}

\noindent and

\begin{equation}
\frac{v_{r,0}v_{\phi,0}}{2}+\nu j_{r,0} = 0\,,
\label{eq:mom2phi0}
\end{equation}

\noindent respectively. Since the magnetic field is vertical at the midplane, $e_{\rm{c}\it{r}\parallel,0} = 0$, $e_{\rm{c}\it{r}\perp,0} = e_{\rm{c}\it{r},0} = e_{r,0} + v_{\phi,0}$, $e_{\rm{c}\it{\phi}\parallel,0} = 0$, and $e_{\rm{c}\it{\phi}\perp,0} = e_{\rm{c}\it{\phi},0} = e_{\phi,0}-v_{r,0} = \left(\epsilon-\epsilon_{\rm B}\right)\theta^{1/2}$.  The radial and azimuthal midplane current densities are therefore $j_{r,0} = -\tilde{\sigma}_{\rm H,0} e_{\rm{c}\it{\phi},0} + \tilde{\sigma}_{\rm P,0}e_{\rm{c}\it{r},0}$ and $j_{\phi,0} = \tilde{\sigma}_{\rm H,0}e_{\rm{c}\it{r},0} + \tilde{\sigma}_{\rm P,0}e_{\rm{c}\it{\phi},0}$ (see Equations (\ref{eq:ohm2r}) and (\ref{eq:ohm2phi})).  Substituting these relations into Equations (\ref{eq:mom2r0}) and (\ref{eq:mom2phi0}) gives
 
\begin{align}
&\frac{\epsilon^{2}\theta}{2}+v_{\phi,0}^{2}-1+\left(4-\frac{2}{\zeta}\right)\theta \nonumber \\
&+\nu \left(\tilde{\sigma}_{\rm H,0}\left[e_{r,0} + v_{\phi,0}\right]+ \tilde{\sigma}_{\rm P,0}\left[\epsilon-\epsilon_{\rm B}\right]\theta^{\frac{1}{2}}\right) = 0 \,,
\label{eq:mom2r02}
\end{align}

\noindent and

\begin{align}
&-\frac{\epsilon\theta^{\frac{1}{2}}v_{\phi,0}}{2} \nonumber \\
& + \nu \left(-\tilde{\sigma}_{\rm H,0} \left[\epsilon-\epsilon_{\rm B}\right]\theta^{\frac{1}{2}} + \tilde{\sigma}_{\rm P,0}\left[e_{r,0} + v_{\phi,0}\right]\right) = 0\,.
\label{eq:mom2phi02}
\end{align}

\noindent Combining these two equations in the unknowns $v_{\phi,0}$ and $e_{r,0}$ gives

\begin{align}
v_{\phi,0}^{2}+v_{\phi,0}\left(\frac{\tilde{\sigma}_{\rm H,0}\epsilon\theta^{\frac{1}{2}}}{2\tilde{\sigma}_{\rm P,0}}\right)+\left[\frac{\epsilon^{2}\theta}{2}-1+\left(4-\frac{2}{\zeta}\right)\theta\right. \nonumber \\
\left.+\nu\left(\frac{\tilde{\sigma}_{\rm H,0}^{2}+\tilde{\sigma}_{\rm P,0}^{2}}{\tilde{\sigma}_{\rm P,0}}\right)\left(\epsilon-\epsilon_{\rm B}\right)\theta^{\frac{1}{2}}\right] = 0 \,,
\label{eq:mom2r03}
\end{align}

\noindent which can be solved for $v_{\phi,0}$.  Either Equation (\ref{eq:mom2r02}) or Equation (\ref{eq:mom2phi02}) can then be used to calculate $e_{r,0}$.

\begin{deluxetable*}{ccccc}
\tablecolumns{5}
\tablewidth{0pc}
\tablecaption{Summary of Midplane Values}
\tablehead{ \multicolumn{1}{c}{Variable} &   \colhead{}   &
\multicolumn{1}{c}{Midplane Value} & \colhead{}   & \multicolumn{1}{c}{$s$-Derivative At Midplane} }
\startdata
$x$&&1&&0 \\
$\tilde{\rho}$&&1&&0 \\
$v_{r}$&&$-\epsilon \theta^{\frac{1}{2}}$&&0 \\
$v_{\phi}$&&$v_{\phi,0}$&&0 \\
$v_{z}$&&0&&$-\left(\frac{5}{2}-\frac{2}{\xi}\right)\epsilon \theta^{\frac{1}{2}}$ \\
$b_{r}$&&0&&$j_{\phi,0}+\left(\frac{1}{\xi}-2\right)$ \\
$b_{\phi}$&&0&&$-j_{r,0}$ \\
$b_{z}$&&1&&0 \\
$e_{r}$&&$\left(\frac{\epsilon\theta^{\frac{1}{2}}}{2\nu\tilde{\sigma}_{\rm P,0}}-1\right)v_{\phi,0}
+\frac{\tilde{\sigma}_{H,0}}{\tilde{\sigma}_{P,0}}\left(\epsilon-\epsilon_{\rm B}\right)\theta^{\frac{1}{2}}$&&0 \\
$e_{\phi}$&&$-\epsilon_{\rm B} \theta^{\frac{1}{2}}$&&0 \\
$j_{r}$&&$\frac{\epsilon \theta^{\frac{1}{2}}v_{\phi,0}}{2\nu}$&&0 \\
$j_{\phi}$&&$\frac{1}{\nu}\left[1-\frac{\epsilon^{2}\theta}{2}-v_{\phi,0}^{2}-\left(4-\frac{2}{\xi}\right)\theta\right]$&&0 \\
\enddata
\label{tab:bcs}
\end{deluxetable*}

The values of several physical quantities and their $s$ derivatives at the midplane as functions of the model parameters and $v_{\phi,0}$ are summarized in Table~\ref{tab:bcs}.

Note that the expression for $v_{z,0}^{\prime}$ is calculated using the mass continuity Equation (\ref{eq:mass2}), rather than Equation (\ref{eq:mom2z}), which is identically 0 at the midplane.  Also, from Equation (\ref{eq:ampz2}) we have $j_{z,0} = 0$.  The magnetic field is vertical at the midplane, so $e_{{\rm c}z,0} = e_{{\rm c}z\parallel,0}$.  From Ohm's law (\ref{eq:ohm2z}) we have $j_{z,0} = \tilde{\sigma}_{\rm O,0} e_{{\rm c}z,0}$; since $\tilde{\sigma}_{\rm O,0}$ is non-zero, $j_{z,0} = 0$ implies $e_{{\rm c}z,0} = 0$.

The midplane is a singular point of the differential equations describing the disk, and special care is needed in initiating the numerical integration from that location.  A common choice for dealing with this issue is to approximate the values of the dependent variables a small distance above the midplane using Taylor series expansions, a task that is made much easier by the assumption that all physical quantities have either even or odd symmetry about the midplane, which implies that the Taylor series for all quantities contain only even power terms or only odd power terms---for instance, the fact that $b_{r,0}^{\prime} \neq 0$ immediately implies that $b_{r,0}^{\prime\prime} = 0$.  Looking at Table~\ref{tab:bcs}, it is clear that second-order Taylor series expansions for $v_{z}$, $b_{r}$ and $b_{\phi}$ only have one term apiece, corresponding to the first derivatives of those quantities at the midplane, which can be calculated using the expressions in Table~\ref{tab:bcs} and Equation (\ref{eq:mom2r03}) for $v_{\phi,0}$.  The second-order Taylor series expansions for $x$, $\tilde{\rho}$, $v_{r}$, $v_{\phi}$, and $e_{r}$ include terms involving $x_{0}^{\prime\prime}$, $\tilde{\rho}_{0}^{\prime\prime}$, $v_{r,0}^{\prime\prime}$, $v_{\phi,0}^{\prime\prime}$, and $e_{r,0}^{\prime\prime}$; the procedure for calculating these quantities is described in Appendix \ref{taylorgen}.

\subsection{Ambipolar Diffusion Case}\label{caseAD}

The ambipolar conductivity regime occurs when all charged species $j$ have $\beta_{j} \gg 1$.  Substituting into Equations (\ref{eq:sigmaohm})--(\ref{eq:sigmapedersen}) gives $\tilde{\sigma}_{\rm O} \gg \tilde{\sigma}_{\rm P} \gg |\tilde{\sigma}_{\rm H}|$ in the ambipolar diffusion limit.

I assume that the charged species consist of ions (subscript `i') and electrons (subscript `e').  Charge neutrality and a comparison of the rate coefficients for electron--neutral and ion--neutral momentum exchange given by \citet{dra83} imply that the heavier ions dominate the
charge--neutral drag force.  Combining this with the balancing of the charge--neutral drag and Lorentz forces yields
\begin{equation} 
\frac{\mbox{\boldmath$J$} \times \mbox{\boldmath$B$}}{c \rho} = \gamma
\rho_{\rm i} \left(\mbox{\boldmath$V$}_{\rm i}-\mbox{\boldmath$V$}\right) \,,
\label{eq:momentad}
\end{equation} 
\noindent where \boldmath$V$\unboldmath$_{\rm i}$ is the ion velocity and $\rho_{\rm i}$ is the ion mass density.  

For a disk in the ambipolar diffusion limit where the only charged species are ions and electrons, the Elsasser number is $\Lambda = \gamma \rho_{\rm i}/\Omega_{\rm K} \equiv \Upsilon$, the ratio of the dynamic timescale to the neutral--ion momentum exchange time \citep{ks11}.  This ratio is denoted by $\eta$ in L96.  The neutral--field coupling parameter $\Upsilon$ in this case is related to the Pedersen conductivity by $\Upsilon = (4\pi v_{\rm A}^{2}/c^{2})(\sigma_{\rm P}/\Omega_{\rm K})$ (SKW07).

The tight coupling of the ions to the magnetic field in the ambipolar diffusion limit implies that the comoving electric field vanishes in the frame of the ions.  The electric field in the disk is therefore
\begin{equation} 
\mbox{\boldmath$E$} = -\frac{1}{c} \mbox{\boldmath$V$}_{\rm i} \times \mbox{\boldmath$B$} \,.
\label{eq:efield} 
\end{equation}

\subsubsection{Governing Equations}\label{goveqad}

Equations (\ref{eq:induct}) and~(\ref{eq:efield}) yield the induction equation
\begin{equation} 
\frac{\partial \mbox{\boldmath$B$}}{\partial t} = \nabla \times
\left(\mbox{\boldmath$V$}_{\rm i} \times \mbox{\boldmath$B$}\right) \,.
\label{eq:induct2} 
\end{equation}

From Equation (\ref{eq:efield}) and the definition of $\epsilon_{\rm B}$, $cE_{\phi} = V_{\rm{i}\mathit{r}}B_{z}-V_{\rm{i}\mathit{z}}B_{r} = -\epsilon_{\rm B}c_{\rm s}(r)B_{z}(r,z=0)$, or

\begin{equation} 
V_{\rm{i}\mathit{z}}B_{r} = V_{\rm{i}\mathit{z}}B_{z} + \epsilon_{\rm B}c_{\rm s}B_{z}(r,z=0) \,,
\label{eq:Virz}
\end{equation}
where all quantities [except the midplane quantity $B_{z}(r,z=0)$] are evaluated at a point $(r,z)$ on the field line of interest.

The $\phi$ component of the induction Equation (\ref{eq:induct}) is
\begin{equation} 
c\left(\frac{\partial E_{z}}{\partial r} - \frac{\partial
E_{r}}{\partial z}\right) = 0 \,, 
\label{eq:inductphi2AD} 
\end{equation}
or
\begin{equation} 
\frac{\partial}{\partial r} \left(V_{\rm{i}\mathit{\phi}}B_{r}-V_{\rm{i}\mathit{r}}B_{\phi}\right) =
\frac{\partial}{\partial z}\left(V_{\rm{i}\mathit{z}}B_{\phi}-V_{\rm{i}\mathit{\phi}}B_{z}\right) \,.  
\label{eq:inductphi3} 
\end{equation}

\subsubsection{Self-Similarity Equations}\label{selfAD}

The ordinary differential equations obtained under the radial self-similarity assumption are identical to Equations (12)--(14) and (16)--(21) in L96.  The equations used here differ from
those in L96 only in the treatment of the induction equation; I allow
for a nonzero $\epsilon_{\rm B}$ (or, equivalently, a nonzero $E_{\phi}$), which
changes the usual flux-freezing condition for the ions.  As a result,
Equation (10) in L96 is replaced by Equations (\ref{eq:Virz})
and (\ref{eq:inductphi3}).  Equations (22)--(24) in L96 (the
self-similar, nondimensionalized versions of Equation (10) in L96) are
replaced by the self-similar, nondimensionalized versions of
Equation (\ref{eq:Virz}),
\begin{equation} 
v_{\rm{i}\mathit{z}}b_{r} = v_{\rm{i}\mathit{r}}b_{z} + \epsilon_{\rm B}\theta^{1/2}x^{1/\zeta-5/2} \,,
\label{eq:virz}
\end{equation}
and~Equation (\ref{eq:inductphi3}),
\begin{align}
 v_{\rm{i}\mathit{r}}^{\prime}b_{\phi} - v_{\rm{i}\mathit{\phi}}^{\prime}b_{r} &-
b_{r}^{\prime}\frac{b_{\phi}}{b_{r}}\left(v_{\rm{i}\mathit{r}} +
\epsilon_{\rm B}\theta^{1/2}x^{1/\zeta-1/2}\right) \nonumber \\
&+ b_{\phi}^{\prime}\left(v_{\rm{i}\mathit{r}} +
\epsilon_{\rm B}\theta^{1/2}x^{1/\zeta-1/2}\right) = \nonumber \\
&b_{r}x^{2}\left(v_{\rm{i}\mathit{r}}b_{\phi} -
v_{\rm{i}\mathit{\phi}}b_{r}\right) \,,
\label{eq:virphiprime}
\end{align}
where the normalization scheme from Section~\ref{selfgen} is used, with the addition of
\begin{displaymath}
\mbox{\boldmath$V$}_{\rm i} = v_{\rm K*} \varphi^{-\frac{\zeta}{2}} \mbox{\boldmath$v$}_{\rm i}\left(s \right) \nonumber \,.
\end{displaymath}

Taking the dot product of \boldmath$B$\unboldmath with both sides of
Equation (\ref{eq:momentad}) gives \boldmath$B$\unboldmath$ \cdot \left(\mbox{\boldmath$V$}_{\rm i} - \mbox{\boldmath$V$}\right) = 0$, which results in the following algebraic relation for
$v_{\rm{i}\mathit{\phi}}$:
\begin{align}
 v_{\rm{i}\mathit{\phi}}b_{\phi} = & b_{r}v_{r} + b_{\phi}v_{\phi} + b_{z}v_{z}
-v_{\rm{i}\mathit{r}}\left(b_{r}+\frac{b_{z}^{2}}{b_{r}}\right) \nonumber \\
&- \epsilon_{\rm B}
\theta^{1/2}\frac{b_{z}}{b_{r}}x^{1/\zeta-5/2} \,,
\label{eq:virphi}
\end{align}
which can be differentiated to give $v_{\rm{i}\mathit{\phi}}^{\prime}$ for use in
Equation (\ref{eq:virphiprime}).

I now have a system of 8 ordinary differential equations in the
variables $x$, $\tilde{\rho}$, $v_{r}$, $v_{\phi}$, $v_{z}$,
$v_{\rm{i}\mathit{r}}$, $b_{r}$ and $b_{\phi}$.  This system differs
from that of L96 in the treatment of the induction equation.  A solution
of this version of the disk equations and a matching wind solution are
presented in Section~\ref{res}.  That solution assumes a functional form for
the matter--field coupling parameter $\Upsilon \propto \tilde \rho_{\rm i}$ that differs from the one used in L96---while L96 assumes that $\Upsilon$ increases monotonically
with height, I take $\Upsilon$ to be constant along a field line, as
would be the case when the ion density is constant (WK93).  In fact, the
coupling parameter is expected to {\em decrease} with height in the
upper reaches of real protostellar disks not too far from the origin
\citep{sw05}, reflecting the decrease of the ion mass density $\tilde\rho_{\rm i}$
with height (which occurs even though the ionization fraction increases
with height because the neutral mass density $\tilde \rho$ drops rapidly
with $z$).

On the other hand, the coupling
parameter increases with radial position because the Keplerian frequency
$\Omega_{\rm K}$ decreases.  Taking $\Upsilon$ constant is a compromise
between these competing effects as the integration proceeds upwards and
outwards along a field line.  Finally, the solution presented here
differs from the solutions presented in L96 in that I initiate the
numerical integration of the disk equations using a Taylor series
expansion (see Appendix \ref{taylorAD}), whereas L96 uses the
approximation $v_{z} = v_{r} s$ in a small region near the midplane to
obtain algebraic relations for $v_{r}$ and $v_{\phi}$.

\subsubsection{Values at the Midplane}\label{midAD}

From Equation (\ref{eq:virz}), $v_{\rm{i}\mathit{r},0} = -\epsilon_{\rm
B} \theta^{1/2}$.  The $r$ component of momentum conservation (Equation
(\ref{eq:moment})), with Equation (\ref{eq:momentad}) used to replace
the Lorentz force term with an ion-neutral drag force term, yields
Equation (13) of L96. At the midplane, this equation reduces to

\begin{equation}
\frac{v_{r,0}^{2}}{2}+v_{\phi,0}^{2}-1+\left(4-\frac{2}{\zeta}\right)\theta+\Upsilon_{0}\left(v_{\rm{i}\mathit{r},0}-v_{r,0}\right) = 0 \,,
\label{eq:vphi0ad}
\end{equation}

\noindent or

\begin{equation}
v_{\phi,0}^{2} = 1 - \frac{\epsilon^{2}\theta}{2} - \left(4-\frac{2}{\zeta}\right)\theta + \Upsilon_{0}\theta^{\frac{1}{2}}\left(\epsilon - \epsilon_{\rm B}\right) \,,
\label{eq:vphi0ad2}
\end{equation}

\noindent which gives $v_{\phi,0}$ in terms of the parameters $\theta$, $\zeta$, $\Upsilon_{0}$, $\epsilon$, and $\epsilon_{\rm B}$.  Continuity of $v_{\rm{i}\mathit{z}}$ implies $v_{\rm{i}\mathit{z},0} = 0$, and the $\phi$ component of the momentum equation (cf. Equation (14) of L96) evaluated at the midplane gives

\begin{equation}
\Upsilon_{0}\left(v_{\rm{i}\mathit{\phi},0} - v_{\phi,0}\right) = \frac{v_{r,0}v_{\phi,0}}{2}\,,
\label{eq:viphi0ad}
\end{equation}

\noindent which can be used to calculate $v_{\rm{i}\mathit{\phi},0}$.

Combining Amp\`{e}re's law (Equation (\ref{eq:amp})) with Equation (\ref{eq:momentad}) yields equations for $b_{r}^{\prime}$ and $b_{\phi}^{\prime}$ (cf. Equations (17)--(19) of L96); evaluating at the midplane yields

\begin{equation}
b_{r,0}^{\prime} = \frac{\Upsilon_{0}}{\nu}\left(v_{\rm{i}\mathit{r},0} - v_{r,0}\right) - \left(2 - \frac{1}{\zeta}\right)
\label{eq:pbr0ad}
\end{equation}

\noindent and

\begin{equation}
b_{\phi,0}^{\prime} = \frac{\Upsilon_{0}}{\nu}\left(v_{\rm{i}\mathit{\phi},0} - v_{\phi,0}\right) \,.
\label{eq:pbphi0ad}
\end{equation}

\noindent The expression for $v_{z,0}^{\prime}$ (see Table~\ref{tab:bcs}) remains unchanged.

The Taylor expansion procedure used to initiate the integration is described in Appendix \ref{taylorAD}.

\section{MODEL PARAMETER CONSTRAINTS}\label{paramconstr}

\subsection{Model Parameters}\label{params}

The dimensionless parameters $\theta \equiv c_{\rm s}^{2}/v_{\rm K}^{2}$
and $\nu \equiv v_{\rm A,0}^{2}/v_{\rm K}^{2}$ are model inputs, as
are the conductivity tensor component profiles $\sigma_{\rm
O}\left(s\right)$, $\sigma_{\rm H}\left(s\right)$, and $\sigma_{\rm
P}\left(s\right)$.  Alternatively, values for $\theta$ and $\nu$
together with a profile $\Upsilon\left(s\right)$ for the neutral--field
coupling parameter suffices for the case of a disk in the ambipolar
diffusion regime.  In either case, a disk/wind solution involves
constraining three parameters to satisfy certain conditions imposed by
physical processes outside the disk.  The first constrained parameter is
$\zeta$, the power-law index of the magnetic flux distribution (Equation (\ref{eq:psidef})).  The
second is $\epsilon$, the normalized midplane value of $V_{r}$ (Equation (\ref{eq:epsdef})). The
third is $\epsilon_{\rm B}$, the normalized midplane value of
$E_{\phi}$ (Equation (\ref{eq:epsbdef})).

Two of the conditions to be satisfied are that the disk solution pass
smoothly through the sonic point and match onto a wind solution that
passes smoothly through the Alfv\'en point.  The third condition relates
the angle of the magnetic field at the disk surface to the distribution
of the magnetic field exterior to the disk (see Section~\ref{surfconstr}).

\subsection{Matching to a Wind Solution}\label{match}

The sonic point ($z=z_{\rm sp}$) is a singular point of the system of
equations describing the disk (see Equation (\ref{eq:rhoprime}) and subsequent discussion),
so it is not possible to integrate directly into the sonic point, but it
is possible to get close.  I run a bisection routine on $\epsilon$,
holding $\zeta$ and $\epsilon_{\rm B}$ fixed, until I obtain a disk
solution that accelerates smoothly almost all the way up to the sonic
point.  I then extrapolate the solution to $z_{\rm sp}$ and attempt to
match the disk solution to a BP82-type wind solution, using the
procedure outlined in SKW11. That is, I calculate the value of
the wind parameter $\kappa$ (the normalized mass-to-magnetic flux ratio)
using the disk solution at the sonic point, and I calculate $B_{r,{\rm
s}}/B_{z,{\rm s}}$ and $\lambda_{\rm s}$, respectively the ratio of the
radial and vertical components of the magnetic field ($\xi_{\rm
b}^\prime$ in the notation of SKW11) and the total specific
angular momentum, using the disk solution at the disk surface.  Here, the disk surface is
taken to be the point where the azimuthal neutral gas velocity matches
the local Keplerian rotation rate.  I run a bisection routine on $\zeta$
until the values of $\kappa$, $\lambda$ and $\xi_{\rm b}^\prime$
calculated from the disk solution match the values at the base of a wind
solution.

The matching procedure used here guarantees continuity of the magnetic
field components, the poloidal mass flux, and the total specific angular
momentum.  However, it should be pointed out that the electric field is,
in general, not continuous at the transition from the disk solution to
the wind solution: in particular, there is a non-zero azimuthal comoving
electric field $E_{\rm{c}\phi}$ in the disk solution because the
parameter $\epsilon_{\rm B}$ is generally non-zero, but the cold, ideal
MHD equations describing the wind solution have $E_{\rm{c}\phi} = 0$
identically.  In addition, the BP82 wind solution is derived for a
razor-thin disk that has vanishing radial and vertical gas velocities
at the base of the wind (negligible compared with the azimuthal
velocity, which is exactly Keplerian at the base of the wind); these are
good approximations when the region of interest is the large-scale
outflow, but break down on small scales comparable to the thickness of
the disk.  The wind solution is therefore expected to give physically
uninteresting results very close to the disk surface.  In fact, the wind
solution has a singular point at the midplane just as the disk solution
does.  I follow BP82 in integrating the wind equations downwards
starting from the Alfv\'en point and stopping very close to the disk
surface, where the solution is matched to a Taylor expansion.  I then
interpolate between the disk solution at the top of the disk and the
wind solution at a distance $\chi/\chi_{\rm A} = 0.002$ above the disk
surface, where $\chi \equiv z/r_{\rm s}$, $r_{\rm s}$ is the radial
distance of the disk surface along the field line of interest, and
$\chi_{A}$ is the value of $\chi$ at the Alfv\'en point.

The wind solution extends past the Alfv\'en point.  There is a second critical point of the equations for a self-similar wind corresponding to the modified fast magnetosonic point (BP82).  However, before a cold BP82-type wind solution reaches the modified fast magnetosonic point, the flow ``over-collimates'' as the field lines bend back towards the axis of symmetry.  This unphysical behavior results from ignoring heating of the jet material and thermal pressure support in the cold wind model.  ``Warm'' self-similar jet models whose solutions do cross the modified fast magnetosonic have been found by  \citet{vl00} and \citet{fc04}.  For this study, I terminate the wind solution when the field lines begin to bend back towards the jet axis.

\subsection{The Disk Surface Field Constraint}\label{surfconstr}

OL01 pointed out that in a self-consistent, global model of a magnetized
disk, the inclination of the poloidal field at the disk surface can be
determined under the approximation that the external magnetic field is
nearly potential ($\nabla \times \mbox{\boldmath$B$} \approx 0$).  An analytic
expression relating $B_{r,{\rm s}}$ to the distribution of $B_{z,{\rm
s}}$ along the disk surface was derived, under this
approximation and assuming an infinitely thin disk, by \citet{cm93}, who
modeled magnetically supported, self-gravitating molecular cloud cores,
and by \citet{lpp94}, who modeled flux advection in viscous accretion
disks. The poloidal field inclination at the disk's surface is directly
related to the balance between the inward advection of magnetic flux by
the accretion flow and the outward diffusion of magnetic flux through
the non-ideal disk. This connection follows from the fact that the radial
component of the magnetic force (per unit area) at the disk surface,
which is $\propto B_{r,{\rm s}} B_{z,{\rm s}}$, is balanced in the ambipolar diffusion regime by the radial ion--neutral drag
force, which is $\propto V_{{\rm i}r} - V_{r}$ (or, taking the midplane
values, $\propto \epsilon - \epsilon_{\rm B}$, which indicates the
explicit relation of $b_{r,{\rm s}}$ to the disk model
parameters). Determining $B_{r,{\rm s}}$ self-consistently in this
manner generally implies a nonzero value of $\epsilon_{\rm B}$.  The
relevant expression for $B_{r,{\rm s}}$ is
\begin{align}
&rB_{r,{\rm s}}(r) = \nonumber \\
&\int_0^{\infty} dk \, k J_{1}(kr)
\int_0^{\infty} dr^{'} r^{'} [B_{z,0}(r^{'})-B_{\rm ref}]
J_{0}(kr^{'}) \,, 
\label{eq:disksurf}
\end{align}
where $J_{0}$ and $J_{1}$ are Bessel functions of the first kind of
order 0 and 1, respectively, $B_{z,0}(r)$ is the midplane vertical field
component at radius $r$, which is assumed to remain constant up to
the surface [so that $B_{z,\rm s}(r)=B_{z,0}(r)$], and $B_{\rm ref}$
is the external magnetic field at ``infinity,'' which is taken to be
$\ll B_{z,0}(r)$, so it may be safely ignored in the second integral.

For a power-law profile $B_{z,0} \propto r^{-a}$,
Equation (\ref{eq:disksurf}) yields $B_{r,{\rm s}}(r)/B_{z,{\rm s}}(r)=
[\Gamma(1-{a}/{2})\, \Gamma({1}/{2}+{a}/{2})][\Gamma({a}/{2})
\, \Gamma({3}/{2} - {a}/{2})]$ for $Re\{a\} \in
(-{3}/{2}, -{1}/{2})$, where $\Gamma$ is the gamma function.  The power-law exponent is related to the disk
model parameter $\zeta$ by $a = 2
- {1}/{\zeta}$, so 
\begin{equation} 
\frac{B_{r,{\rm s}}(r)}{B_{z,{\rm s}}(r)}= \frac{\Gamma\left[1/\left(2\zeta\right)\right]\,
\Gamma\left[3/2-1/\left(2\zeta\right)\right]}{\Gamma\left[1-1/\left(2\zeta\right)
\right] \, \Gamma\left[1/2 + 1/\left(2\zeta\right)\right]} \;.
\label{eq:disksurf2}
\end{equation}
This result holds for $\zeta \in ({2}/{3}, 2)$ and is therefore
appropriate for the values $\zeta \lesssim {4}/{3}$ that are expected to
characterize physically viable solutions (e.g., L96; F97).  For $\zeta = {4}/{3}$, Equation (\ref{eq:disksurf2}) yields $B_{r,{\rm s}}(r)/B_{z,{\rm s}}(r) = 1.428$.

Once a disk solution that passes smoothly through the sonic point has
been matched to a wind solution, I bisect on $\epsilon_{\rm B}$ until
the value of $B_{r, \rm s}$ matches the value imposed by
the field distribution.

\section{ILLUSTRATIVE SOLUTION}\label{res}

\begin{figure}[t]
\epsscale{1.0}
		\plotone{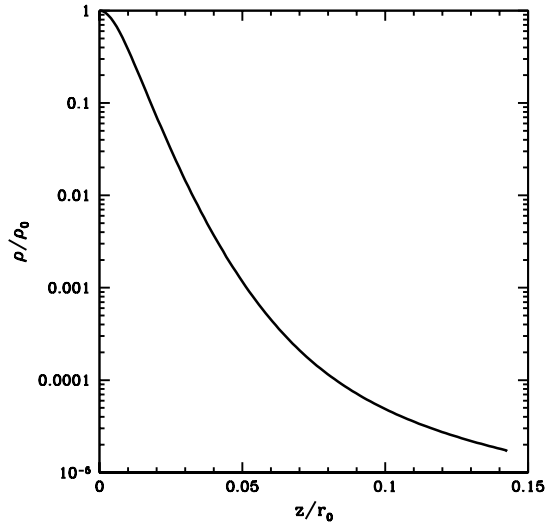}
		\caption{Density structure of the constant $\Upsilon$ disk solution.}
		\label{fig:disk1rho}
\end{figure}

\begin{figure}[ht]
\epsscale{1.0}
		\plotone{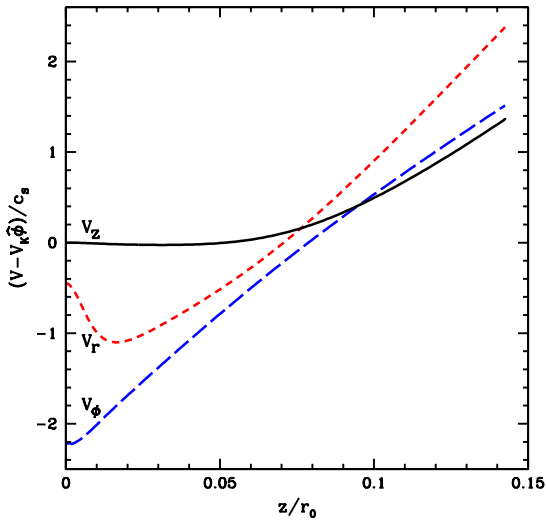}
		\caption{Neutral velocity structure of the constant $\Upsilon$ disk solution.  The Keplerian velocity has been subtracted from the azimuthal component, and the plotted velocities are normalized by the local sound speed.}
		\label{fig:disk1v}
\end{figure}

\begin{figure}[ht]
\epsscale{1.0}
		\plotone{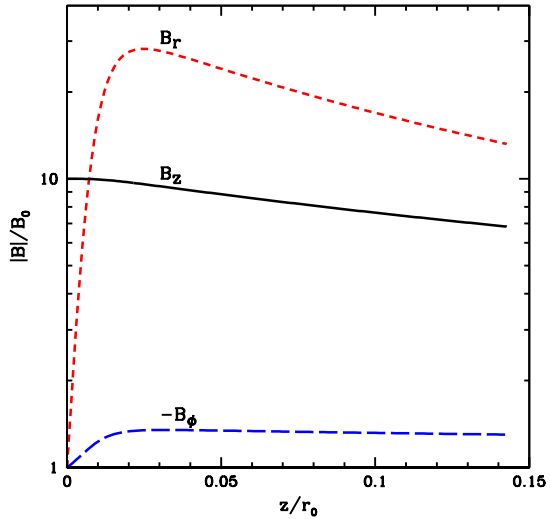}
		\caption{Magnetic field structure of the constant $\Upsilon$ disk solution.  All components are normalized by the midplane field strength.}
		\label{fig:disk1b}
\end{figure}

\begin{figure}[ht]
\epsscale{1.0}
		\plotone{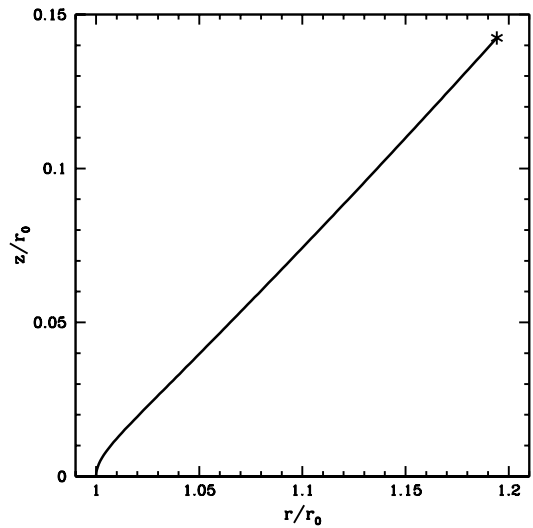}
		\caption{Poloidal magnetic field line shape of the constant $\Upsilon$ disk solution.  The location of the sonic point is marked (`*').}
		\label{fig:disk1fieldline}
\end{figure}

Protostellar disks have typical inferred temperatures at 1 AU of $\sim
100-300$ K, masses of $\sim 0.005-0.2 M_{\odot}$, column densities at 5
AU of $\sim 0.3-300$ g cm$^{-2}$, and extend to radii $\sim 200$ AU
\citep[e.g.,][]{aw07}.  Disk/wind model solutions typically have $v_{\rm
A,0}/c_{\rm s} \lesssim 1$ (e.g., WK93), and observations of protostellar
systems with jets indicate that the ratio of the outflow and accretion
rates is $\sim 0.01-0.1$ \citep[e.g.,][]{har95}.  Finally, observations
of remanent magnetization of meteorite samples suggest a magnetic field
of $\sim 1$ G at 3 AU in the early solar nebula \citep{ls78}.  Motivated
by these facts, I choose parameter values $\theta = 1 \times 10^{-3}$
and $\nu = 7.744 \times 10^{-4}$.  These values together with the
constant coupling profile $\Upsilon\left(s\right) = 10$ yield a solution
of the disk equations described in Section~\ref{caseAD}. This solution passes smoothly
through the sonic point, matches to a wind solution that passes smoothly
through the Alfv\'en point, and satisfies the constraint on the magnetic
field at the surface of the disk.  The corresponding disk and wind
parameters are $\delta\zeta = 3.05 \times 10^{-3}$, $\epsilon = 0.45$,
$\epsilon_{\rm B} = 0.03$, $\kappa = 1.27 \times 10^{-3}$, $\lambda =
108.4$, and $\xi_{\rm b}^\prime = 1.42$. The small value of
$\delta\zeta$ found here is consistent with the predictions of L96 and
F97 that cold wind-driving disks should have field-line distributions
very close to the limiting case of $\zeta = 4/3$ (but see \citealt{cf00} for arguments that larger values of $\delta\zeta$ can occur in
wind-driving disks when thermal effects are considered\footnote{The
quantity $\delta\zeta$ is related to the quantity $\xi$ in \citet{cf00}
as $\delta\zeta \approx 8/9~\xi$; the ratio $f$ of the mass outflow in the
wind between radii $r_{\rm in}$ and $r_{\rm max}$ to the mass accretion rate at
$r_{\rm in}$ is $f = (r_{\rm max}/r_{\rm in})^{\xi}-1$.  For $(r_{\rm max}/r_{\rm in}) \approx
1000$ and an observed $f \sim 0.01-0.1$, $\xi$ is inferred to lie in
the range $\xi \sim 0.0014-0.014$.  Of the two representative solutions
in \citet{cf00}, the one with almost no entropy generation gives $\xi
\sim 0.001$, while the one with strong entropy generation gives $\xi
\sim 0.5$, in which case almost all of the disk mass flows out
into the wind rather than accreting onto the central protostar.}).

\begin{figure}[ht]
\epsscale{1.0}
		\plotone{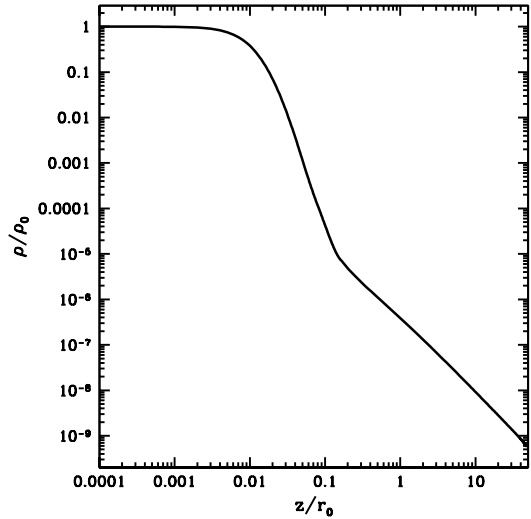}
		\caption{Density structure of the constant $\Upsilon$ disk solution and the matched wind solution.}
		\label{fig:diskwind1rho}
\end{figure}

\begin{figure}[ht]
\epsscale{1.0}
		\plotone{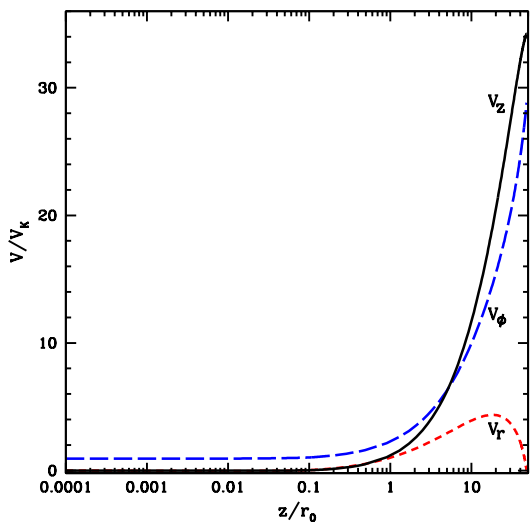}
		\caption{Velocity structure of the constant $\Upsilon$ disk solution and the matched wind solution.  The plotted velocities are normalized by the local Keplerian velocity.}
		\label{fig:diskwind1v}
\end{figure}

\begin{figure}[ht]
\epsscale{1.0}
		\plotone{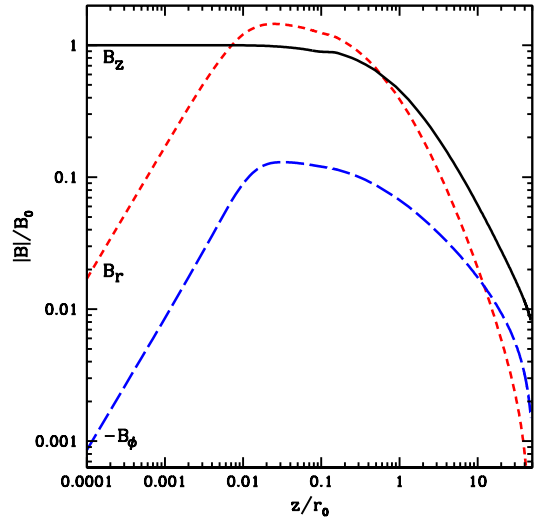}
		\caption{Magnetic field structure of the constant $\Upsilon$ disk solution and the matched wind solution.  All components are normalized by the midplane field strength.  The radial and toroidal components are 0 at the midplane, and the vertical component is 1 at the midplane, but all three components change rapidly within the disk, which extends up to $z/r_{0} \approx 0.079$.}
		\label{fig:diskwind1b}
\end{figure}

\begin{figure}[ht]
\epsscale{1.0}
		\plotone{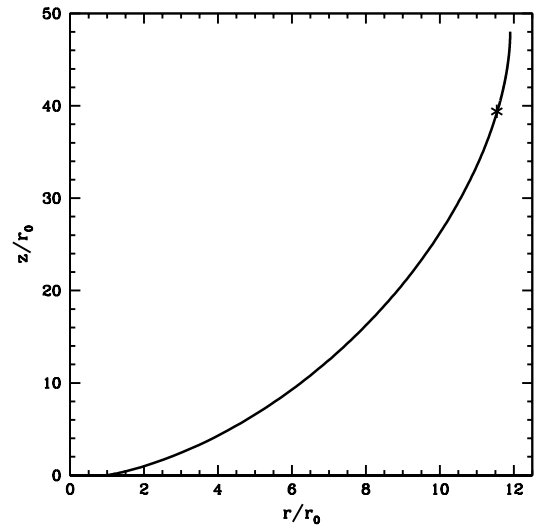}
		\caption{Poloidal magnetic field line shape of the constant $\Upsilon$ disk solution and the matched wind solution.  The location of the Alfv\'en point is marked (`*').}
		\label{fig:diskwind1fieldline}
\end{figure}

A rough estimate of the ratio of the mass outflow rate from both sides
of the disk, between an outermost radius $r_{\rm max}$ and a given radius
$r$, to the mass accretion rate through the disk at $r$, can be derived
from Equation (89) of SKW11 as $\dot M_{\rm wind}/\dot M_{\rm disk}
\approx \rm{ln}\mathit{(r_{\rm{max}}/r)}/[2(\lambda-1)]$; for the
derived solution, and taking $r_{\rm{max}}/r \approx 1000$, this ratio is
$\approx 3.2\times 10^{-2}$.  For a central protostar of mass 0.5
$M_{\odot}$, the model has a column density of $\approx 42$ g cm$^{-2}$ at 1 AU and a column density of $\approx 19$ g cm$^{-2}$ at 5 AU if the 
density is normalized so that the accretion rate is $\approx 3 \times
10^{-6} \, M_{\odot} \, \rm{yr}^{-1}$ and the total disk mass out to 200
AU is $\approx 0.055 \, M_{\odot}$. In comparison, the minimum mass solar
nebula model \citep{hay81} has a column density at 1 AU of $\approx 1700$ g
cm$^{-2}$ and at 5 AU of $\approx 150$ g cm$^{-2}$.  The density
normalization used here corresponds to a midplane magnetic field
strength at 1 AU of $|\mbox{\boldmath$B$}| \approx 2.5$ G and at 3 AU of $\approx 0.6$
G.  The disk temperature at 1 AU is $\approx$ 125 K.

The gas density, velocity, and magnetic field structures of the disk
solution as functions of position along a field line are shown in
Figures~\ref{fig:disk1rho}--\ref{fig:disk1b}.  The plots extend from
the midplane, $z=0$, to the sonic point, $z_{\rm sp} \approx 0.14\, r_{0} \left(\approx 3.8 \, h_{\rm
T}\right)$. The location of the sonic point is marked in
Figure~\ref{fig:disk1fieldline}, which shows the poloidal magnetic field
line structure in the disk.  The magnetic field components and the
density are normalized by the midplane values of $B$ and $\rho$,
respectively.  The velocity components, with the Keplerian velocity
subtracted from the azimuthal neutral and ion velocities, are normalized
by the local value of the sound speed along the field line (that is,
taking into account the spatial variation of the sound speed).

At the midplane, the field is vertical by symmetry; above the
midplane, the radial component grows rapidly, a small negative toroidal
component develops, and the vertical component decreases slightly.  The
vertical velocity $V_{z}$ starts at zero at the midplane, decreases to
small negative values in the near-midplane region (reflecting the radial
convergence of the accretion flow), and then increases back
through zero at $z \approx 0.054\, r_{0} \left(\approx 1.5\, h_{\rm T}\right)$, becoming comparable to
$c_{\rm s}$ at $z \approx 0.112\, r_{0} \left(\approx 3.1\, h_{\rm T}\right)$. The radial velocity starts out at
$-0.45\,c_{\rm s}$ at the midplane and decreases in the near-midplane
region, reaching a minimum at $z \approx 0.016\, r_{0} \left(\approx 0.5 \, h_{\rm T}\right)$ and then
increasing, switching from an inflow to an outflow at $z \approx 0.071\, r_{0} \left(\approx 2\,
h_{\rm T}\right)$. The azimuthal velocity is sub-Keplerian at the midplane
($\approx 0.93\,V_{\rm K,0}$) but increases smoothly with height, becoming
super-Keplerian at the disk surface, $z_{\rm s} \approx 0.77\, r_{0} \left(\approx 2.2\, h_{\rm T}\right)$, slightly above the point
where $V_{r}$ becomes positive.  In the hydrostatic approximation ($V_{z} = 0$), $V_{r} = 0$ and $V_{\phi} = V_{\rm K}$ at the same point (WK93).  The effective density scale height $h_{\rm eff}$ (defined so that $\rho\left(z = h_{\rm eff}\right)/\rho\left(z = 0\right) = 1/\sqrt{e}$) is $h_{\rm eff} \approx 0.22 \,h_{\rm T}$, indicating that magnetic squeezing has a significant effect on the disk structure (WK93).

The gas density, velocity, and magnetic field structures of the matched
disk and wind solutions as functions of position along a field line
are shown in Figures~\ref{fig:diskwind1rho}--\ref{fig:diskwind1b}.  The
plots extend from $z=0$ to $z \approx 48\, r_{0}$, where the field lines begin to bend back towards the jet axis and the wind solution terminates.  Along the way, the wind solution passes through the Alfv\'en point, located at $z \approx
38.9\, r_{0}$.  Figures~\ref{fig:diskwind1rho}--\ref{fig:diskwind1b} are shown with the variable $z/r_{0}$ plotted on a logarithmic axis so that the structures of the disk and wind regions can be distinguished.  The location of the Alfv\'en point is marked in
Figure~\ref{fig:diskwind1fieldline}, which shows the poloidal magnetic
field line structure in the disk and wind.  In the region above the
disk, the poloidal field quickly drops while the azimuthal field
amplitude first grows slightly ($B_{\phi}$ becomes more negative) and
then decreases, in accordance with the expected behavior of a
wind-driving disk.  The vertical and azimuthal velocities increase
smoothly throughout the outflow, while the radial velocity reaches a
maximum and begins decreasing well below the Alfv\'en point.  Note that
Figure~\ref{fig:diskwind1v} differs from Figure~\ref{fig:disk1v} in that
the unmodified azimuthal velocity is plotted and the velocities are
normalized by the local value of the Keplerian speed (that is, taking
into account the spatial variation of the Keplerian speed).

The solution presented here is similar to the WK93 model in that it considers a strongly coupled, ambipolar diffusive disk, but differs in that it is a global model with a non-zero value of $\epsilon_{\rm B}$ determined self-consistently from a constraint on the field line inclination at the disk surface.  The equilibrium solution curves for the present model are undoubtedly modified from those of K04, but on the assumption that the changes are not drastic, a comparison of the representative solution and the results given in K04 may be instructive.  The representative solution given here has $\kappa = 1.27 \times 10^{-3}$, $\lambda = 108.4$, and $a \equiv \left(\sigma/\theta\right)^{1/2} = 0.88$; the quantity $\eta$ in K04 corresponds to $\Upsilon$.  Comparing these values to Figures 1 and 2 of K04, it appears that the representative solution given here lies near the lowest point of the lower branch of the $\eta = 10$ equilibrium solution curve in Figure 1, and to the left of the upper branch of the $\eta = 10$ equilibrium solution curve in Figure 2.  In both cases, the comparison suggests that the representative solution given here is stable.

This illustrative solution assumes that the underlying
disk is in the ambipolar conductivity regime.  Ambipolar diffusion is
likely a reasonable assumption in the outer regions of typical
protostellar disks, but may not hold throughout the inner regions of
protostellar disks, which are more likely in the Hall or Ohm
conductivity regimes.  However, it is worth noting in this connection
that models of wind-driving disks have column densities in the inner
regions that are much smaller than in typical turbulent-viscosity
accretion disks.  These smaller column densities suggest that the region
of a wind-driving disk in which ambipolar diffusivity is a good
approximation may be substantially larger than in other types of disk
models.

\section{CONCLUSION}\label{concs}
The model for the global structure of a magnetized protostellar disk/wind
system presented here improves on previous studies (e.g.,
WK93; L96; F97; OL01) by including, for the first time, a fully
self-consistent treatment of flux migration on the accretion timescale.
This treatment employs the approximation that the magnetic field on
scales $z \lesssim r$ above the disk surface is very nearly
potential, leading to a simple constraint on the surface field inclination (Equation (\ref{eq:disksurf})).  The model also improves on previous studies (e.g., SKW07) by extending the conductivity tensor formalism for describing non-ideal disk material from radially localized disk models to global, radially self-similar disk models.

The model presented here describes solutions for a disk and wind that are steady on timescales up to the flux migration timescale.  Studies of simplified equilibrium models of disk winds by \citet{lu94}, \citet{cs02}, and \citet{ca09} have raised the possibility that magnetocentrifugal disk winds are inherently unstable.  \citet[][hereafter K04]{k04} showed that for strongly coupled wind-driving disk models of the sort described in WK93 there are two equilibrium solution branches, one stable and one unstable, and argued that protostellar systems correspond to the stable branch.  This line of argument has received some support from simulations of axisymmetric, diffusive accretion disks launching self-collimating, magnetocentrifugal disk winds that have been carried out by \citet{ck02, ck04} and \citet{za07}.  These simulations confirm that the resulting large-scale jets are continuous and steady over many dynamic timescales, providing support for the present model's quasi-steady state assumption.  In turn, the semianalytic model presented here facilitates calculations that cannot currently be carried out in full numerical simulations.

The formulation presented here allows construction of global disk/wind
models with realistic conductivity profiles in the disk.  Finding
matched disk/wind solutions of the sort presented here is a difficult
task due to the imposed restrictions of requiring all angular momentum
transport to occur via the wind, and the strict self-similarity of the
disk and wind solutions.  In reality, protostellar disk/wind systems may
transport angular momentum simultaneously through multiple mechanisms,
such as magnetic braking, gravitational instability, and turbulence 
(see SKW07 for a radially localized disk model that includes angular
momentum transport by both a wind and turbulence driven by the
magnetorotational instability), and are unlikely to be globally
self-similar.  However, simple models of the sort presented here are
valuable tools for investigating global processes in disk/wind systems,
such as dust transport and processing (S. Teitler, in preparation).
Studies of dust evolution processes in wind-driving disks will
complement similar studies carried out in simple models of turbulent
disks \citep[e.g.,][]{kg04, bra08} and in the X-wind model
\citep[e.g.,][]{shu01}, and are of particular importance for
wind-driving disks since small dust grains can have strong effects on
the coupling of neutral disk material to magnetic fields \citep[e.g.,][]{nnu91,w07}.

\bigskip
\bigskip
\acknowledgments
I thank Arieh K\"onigl for his unwearying support and guidance over the past several years, and my thesis committee members Fausto Cattaneo, Fred Ciesla, Joshua Frieman, and Don York for their patience and advice.  I also thank Raquel Salmeron, Konstantinos Tassis, and Mark Wardle for many helpful discussions.  In addition, Raquel Salmeron and Zhi-Yun Li graciously made available unpublished material that proved very helpful.  Finally, I thank the anonymous referee for many helpful comments.  This work was supported in part by NASA Astrophysics Theory and Fundamental Physics Program grant NNX09AH38G and NSF grant AST-0908184.

\newpage

\appendix
\section{Calculating $e_{r}^{\prime}$}\label{appez}
Given the values of the independent variable $s$ and the dependent variables $x$, $v_{r}$, $v_{\phi}$, $v_{z}$, $b_{r}$, $b_{\phi}$, $e_{r}$, and $\tilde{\rho}$, the auxiliary variables $b_{z}$ and $e_{\phi}$ can be obtained directly from the algebraic relations (\ref{eq:brbz}) and (\ref{eq:induct2b}).  Their derivatives are 
\begin{displaymath}
b_{z}^{\prime} = b_{r}^{\prime}s+b_{r}-2\frac{x^{\prime}}{x^{3}} = b_{r}^{\prime}s-b_{r}
\end{displaymath}
\noindent and 
\begin{displaymath}
e_{\phi}^{\prime} = \left(\frac{1}{\zeta}-\frac{5}{2}\right)e_{\phi}\left(0\right)b_{r}x^{\frac{1}{\zeta}-\frac{1}{2}} \nonumber \,. 
\end{displaymath}
The auxiliary variables $e_{\rm{c}\it{r}}$ and $e_{\rm{c}\it{\phi}}$ are obtained from Equation (\ref{eq:ec2}).

The next step is to combine the algebraic relations (\ref{eq:jrjz}), (\ref{eq:ohm2r}), and (\ref{eq:ohm2z}) to obtain

\begin{eqnarray}
\tilde{\sigma}_{\rm O} e_{\rm{c}\it{z}\parallel} + \frac{\tilde{\sigma}_{\rm H}}{b}\left(b_{r}e_{\rm{c}\it{\phi}\perp}-b_{\phi}e_{\rm{c}\it{r}\perp}\right) + \tilde{\sigma}_{\rm P} e_{\rm{c}\it{z}\perp} & = &\left(\tilde{\sigma}_{\rm O} e_{\rm{c}\it{r}\parallel} + \frac{\tilde{\sigma}_{\rm H}}{b}\left(b_{\phi}e_{\rm{c}\it{z}\perp}-b_{z}e_{\rm{c}\it{\phi}\perp}\right) + \tilde{\sigma}_{\rm P} e_{\rm{c}\it{r}\perp}\right)s \nonumber \\
& & + \left(\frac{1}{\zeta}-1\right)\frac{b_{\phi}}{x} \,.
\label{eq:jrjz2}
\end{eqnarray}

\noindent Using Equations (\ref{eq:ecpar}) and (\ref{eq:ecperp}), we have

\begin{eqnarray}
e_{\rm{c}\it{r}\parallel} = e_{\rm{c}\it{r}} - e_{\rm{c}\it{r}\perp}; \qquad \qquad \quad \; \; \, e_{\rm{c}\it{\phi}\parallel} = \frac{b_{\phi}}{b_{r}}\left(e_{\rm{c}\it{r}} - e_{\rm{c}\it{r}\perp}\right); \nonumber &\\
e_{\rm{c}\it{z}\parallel} = \frac{b_{z}}{b_{r}}\left(e_{\rm{c}\it{r}} - e_{\rm{c}\it{r}\perp}\right); \qquad \quad e_{\rm{c}\it{\phi}\perp} = e_{\rm{c}\it{\phi}}-\frac{b_{\phi}}{b_{r}}\left(e_{\rm{c}\it{r}} - e_{\rm{c}\it{r}\perp}\right); \nonumber &\\
e_{\rm{c}\it{z}\perp} = -\frac{1}{b_{z}}\left(b_{r}e_{\rm{c}\it{r}\perp}+b_{\phi}\left[e_{\rm{c}\it{\phi}}-\frac{b_{\phi}}{b_{r}}\left(e_{\rm{c}\it{r}} - e_{\rm{c}\it{r}\perp}\right)\right]\right) \,,
\label{eq:ecrels}
\end{eqnarray}

\noindent giving \boldmath$e$\unboldmath$_{\rm{c}\parallel}$ and \boldmath$e$\unboldmath$_{\rm{c}\perp}$ in terms of ${e_{\rm{c}\it{r}\perp}}$ and the known quantities $e_{\rm{c}\it{r}}$, $e_{\rm{c}\it{\phi}}$, and \boldmath$b$\unboldmath.  Substituting these expressions into Equation (\ref{eq:jrjz2}) and solving, we obtain

\begin{eqnarray}
e_{\rm{c}\it{r}\perp}\left[\tilde{\sigma}_{\rm O}\left(s-\frac{b_{z}}{b_{r}}\right)-\tilde{\sigma}_{\rm P}\left(\frac{b_{r}}{b_{z}}+\frac{b_{\phi}^{2}}{b_{r}b_{z}}+s\right)+s\frac{\tilde{\sigma}_{\rm H}}{b}\left(\frac{b_{r}}{b_{\phi}}{b_{z}}+{b_{\phi}^{3}}{b_{r}b_{z}}+\frac{b_{\phi}b_{z}}{b_{z}}\right)\right] = & \nonumber \\
e_{\rm{c}\it{r}}\left[\tilde{\sigma}_{\rm O}\left(s-\frac{b_{z}}{b_{r}}\right)-\tilde{\sigma}_{\rm P}\frac{b_{\phi}^{2}}{b_{r}b_{z}}+\frac{\tilde{\sigma}_{\rm H}}{b}\left(b_{\phi}+s\left[\frac{b_{\phi}^{3}}{b_{r}b_{z}}+\frac{b_{\phi}b_{z}}{b_{r}}\right]\right)\right] & \nonumber\\
+ e_{\rm{c}\it{\phi}}\left[\tilde{\sigma}_{\rm P}\frac{b_{\phi}}{b_{z}}-\frac{\tilde{\sigma}_{\rm H}}{b}\left(s\left[\frac{b_{\phi}^{2}}{b_{z}}+b_{z}\right]+b_{r}\right)\right] + \left(\frac{1}{\zeta}-1\right)\frac{b_{\phi}}{x} & \nonumber \,,
\end{eqnarray}

\noindent which simplifies to

\begin{equation}
e_{\rm{c}\it{r}\perp} = Ae_{\rm{c}\it{r}} + B = Ae_{r}+B+A\left(v_{r}b_{\phi}-v_{\phi}b_{r}\right)\,,
\label{eq:ecperpr2}
\end{equation}

\noindent where
\begin{displaymath}
A = \frac{\left[\tilde{\sigma}_{\rm O}bb_{z}+\tilde{\sigma}_{\rm P}x^{2}bb_{\phi}^{2}-s\tilde{\sigma}_{\rm H}x^{2}b^{2}b_{\phi}-\tilde{\sigma}_{\rm H}b_{r}b_{\phi}\right]}{\left[\left(\tilde{\sigma}_{\rm O}-\tilde{\sigma}_{\rm P}\right)bb_{z}+\tilde{\sigma}_{\rm P}x^{2}b^{3}-s\tilde{\sigma}_{\rm H}x^{2}b^{2}b_{\phi}\right]} \nonumber
\end{displaymath}

\noindent and
\begin{displaymath}
B = \frac{\left[e_{\rm{c}\it{\phi}}\left(-\tilde{\sigma}_{\rm P}x^{2}bb_{r}b_{\phi}+s\tilde{\sigma}_{\rm H}x^{2}b^{2}b_{r}+\tilde{\sigma}_{\rm H}b_{r}^{2}\right)-\left(\frac{1}{\zeta}-1\right)xbb_{r}b_{\phi}b_{z}\right]}{\left[\left(\tilde{\sigma}_{\rm O}-\tilde{\sigma}_{\rm P}\right)bb_{z}+\tilde{\sigma}_{\rm P}x^{2}b^{3}-s\tilde{\sigma}_{\rm H}x^{2}b^{2}b_{\phi}\right]} \nonumber \,.
\end{displaymath}

\noindent Equation (\ref{eq:ecperpr2}) for $e_{\rm{c}\it{r}\perp}$ in terms of known quantities can be used to infer $e_{\rm{c}\it{r}\parallel}$, $e_{\rm{c}\it{\phi}\parallel}$, $e_{\rm{c}\it{\phi}\perp}$, $e_{\rm{c}\it{z}\parallel}$, $e_{\rm{c}\it{z}\perp}$, $e_{\rm{c}\it{z}}$, $e_{z}$, $j_{r}$, $j_{\phi}$, and $j_{z}$ using Equations (\ref{eq:ecrels}) and (\ref{eq:ohm2r})--(\ref{eq:ohm2z}).  These auxiliary quantities can in turn be used in Equations (\ref{eq:rhoprime}), (\ref{eq:mom2r})--(\ref{eq:mom2z}), (\ref{eq:ampphi2}) and (\ref{eq:ampr2}) to obtain $\tilde{\rho}^{\prime}$, $v_{r}^{\prime}$, $v_{\phi}^{\prime}$, $v_{z}^{\prime}$, $b_{r}^{\prime}$, and $b_{\phi}^{\prime}$.

The differential Equation (\ref{eq:induct2a}) for $e_{r}^{\prime}$ involves the derivative of the auxiliary quantity $e_{z}$; $e_{z}$ is given by

\begin{eqnarray}
e_{z} & = &e_{\rm{c}\it{z}}+v_{\phi}b_{r}-v_{r}b_{\phi} \nonumber \\
& = &e_{\rm{c}\it{z}\parallel}+e_{\rm{c}\it{z}\perp}+v_{\phi}b_{r}-v_{r}b_{\phi} \nonumber \\
& = &\frac{b_{z}}{b_{r}}\left(e_{\rm{c}\it{r}} - e_{\rm{c}\it{r}\perp}\right)-\frac{1}{b_{z}}\left(b_{r}e_{\rm{c}\it{r}\perp}+b_{\phi}\left[e_{\rm{c}\it{\phi}}-\frac{b_{\phi}}{b_{r}}\left(e_{\rm{c}\it{r}} - e_{\rm{c}\it{r}\perp}\right)\right]\right)+v_{\phi}b_{r}-v_{r}b_{\phi} \nonumber \,.
\end{eqnarray}

\noindent Upon substituting for $e_{\rm{c}\it{r}\perp}$ from Equation (\ref{eq:ecperpr2}), this becomes

\begin{eqnarray}
e_{z} & = &e_{r}\left(\frac{b_{z}}{b_{r}}\left[1-A\right]-\frac{b_{r}}{b_{z}}A+\frac{b_{\phi}^{2}}{b_{r}b_{z}}\left[1-A\right]\right) \nonumber \\
& & +\left(v_{\phi}b_{z}-v_{z}b_{\phi}\right)\left(\frac{b_{z}}{b_{r}}\left[1-A\right]-\frac{b_{r}}{b_{z}}A+\frac{b_{\phi}^{2}}{b_{r}b_{z}}\left[1-A\right]\right) \nonumber \\
& & -B\frac{b^{2}}{b_{r}b_{z}}-\frac{b_{\phi}}{b_{z}}e_{\rm{c}\it{\phi}}+v_{\phi}b_{r}-v_{r}b_{\phi} \,.
\label{eq:ezer}
\end{eqnarray}

\noindent Differentiating Equation (\ref{eq:ezer}) for $e_{z}$ gives

\begin{equation}
e_{z}^{\prime} = Ce_{r}^{\prime}+D\,,
\label{eq:ezerprime}
\end{equation}

\noindent where

\begin{displaymath}
C = \frac{b_{z}}{b_{r}}\left(1-A\right)-\frac{b_{r}}{b_{z}}A+\frac{b_{\phi}^{2}}{b_{r}b_{z}}\left(1-A\right) = \frac{b_{z}}{b_{r}}+\frac{b_{\phi}^{2}}{b_{r}b_{z}}-\frac{b^{2}}{b_{r}b_{z}}A \nonumber \,,
\end{displaymath}

\bigskip

\begin{eqnarray}
D & = &C^{\prime}e_{r}+C\left(v_{\phi}^{\prime}b_{z}+v_{\phi}b_{z}^{\prime}-v_{z}^{\prime}b_{\phi}-v_{z}b_{\phi}^{\prime}\right)+C^{\prime}\left(v_{\phi}b_{z}-v_{z}b_{\phi}\right) \nonumber \\
& & -B^{\prime}\frac{b^{2}}{b_{r}b_{z}}-2B\frac{bb^{\prime}}{b_{r}b_{z}}+B\frac{b^{2}b_{r}^{\prime}}{b_{r}^{2}b_{z}}+B\frac{b^{2}b_{z}^{\prime}}{b_{r}b_{z}^{2}}-\frac{b_{\phi}^{\prime}}{b_{z}}e_{\rm{c}\it{\phi}}+\frac{b_{\phi}b_{z}^{\prime}}{b_{z}^{2}}e_{\rm{c}\it{\phi}} \nonumber \\
& & -\frac{b_{\phi}}{b_{z}}e_{\rm{c}\it{\phi}}^{\prime}+\left(v_{\phi}^{\prime}b_{r}+v_{\phi}b_{r}^{\prime}-v_{r}^{\prime}b_{\phi}-v_{r}b_{\phi}^{\prime}\right) \nonumber \,,
\end{eqnarray}

\bigskip

\begin{displaymath}
C^{\prime} = \frac{b_{z}^{\prime}}{b_{r}}-\frac{b_{z}b_{r}^{\prime}}{b_{r}^{2}}+2\frac{b_{\phi}b_{\phi}^{\prime}}{b_{r}b_{z}}-\frac{b_{\phi}^{2}b_{r}^{\prime}}{b_{r}^{2}b_{z}}-\frac{b_{\phi}^{2}b_{z}^{\prime}}{b_{r}b_{z}^{2}}-2\frac{bb^{\prime}}{b_{r}b_{z}}A+\frac{b^{2}b_{r}^{\prime}}{b_{r}^{2}b_{z}}A+\frac{b^{2}b_{z}^{\prime}}{b_{r}b_{z}^{2}}A-\frac{b^{2}}{b_{r}b_{z}}A^{\prime} \nonumber \,, \\
\end{displaymath}

\bigskip

\begin{eqnarray}
A^{\prime} & = &\bigg(\left[\tilde{\sigma}_{\rm O}^{\prime}bb_{z}+\tilde{\sigma}_{\rm O}b^{\prime}b_{z}+\tilde{\sigma}_{\rm O}bb_{z}^{\prime}+\tilde{\sigma}_{\rm P}^{\prime}x^{2}bb_{\phi}+2\tilde{\sigma}_{\rm P}xx^{\prime}bb_{\phi}^{2}+\tilde{\sigma}_{\rm P}x^{2}b^{\prime}b_{\phi}^{2}+2\tilde{\sigma}_{\rm P}x^{2}bb_{\phi}b_{\phi}^{\prime} \right. \nonumber \\
& & \left. -\tilde{\sigma}_{\rm H}x^{2}b^{2}b_{\phi}-s\tilde{\sigma}_{\rm H}^{\prime}x^{2}b^{2}b_{\phi}-2s\tilde{\sigma}_{\rm H}xx^{\prime}b^{2}b_{\phi}^{2}-2s\tilde{\sigma}_{\rm H}x^{2}bb^{\prime}b_{\phi}-s\tilde{\sigma}_{\rm H}x^{2}b^{2}b_{\phi}^{\prime}\right. \nonumber \\
& & \left. -\tilde{\sigma}_{\rm H}^{\prime}b_{r}b_{\phi}-\tilde{\sigma}_{\rm H}b_{r}^{\prime}b_{\phi}-\tilde{\sigma}_{\rm H}b_{r}b_{\phi}^{\prime}\right]\bigg)\bigg/\left[\left(\tilde{\sigma}_{\rm O}-\tilde{\sigma}_{\rm P}\right)bb_{z}+\tilde{\sigma}_{\rm P}x^{2}b^{3}-s\tilde{\sigma}_{\rm H}x^{2}b^{2}b_{\phi}\right] \nonumber\\
& & -\bigg(\left[\tilde{\sigma}_{\rm O}bb_{z}+\tilde{\sigma}_{\rm P}x^{2}bb_{\phi}^{2}-s\tilde{\sigma}_{\rm H}x^{2}b^{2}b_{\phi}-\tilde{\sigma}_{\rm H}b_{r}b_{\phi}\right] \big[\left(\tilde{\sigma}_{\rm O}^{\prime}-\tilde{\sigma}_{\rm P}^{\prime}\right)bb_{z}+\left(\tilde{\sigma}_{\rm O}-\tilde{\sigma}_{\rm P}\right)b^{\prime}b_{z} \nonumber \\
& & +\left(\tilde{\sigma}_{\rm O}-\tilde{\sigma}_{\rm P}\right)bb_{z}^{\prime}+\tilde{\sigma}_{\rm P}x^{2}b^{3}+2\tilde{\sigma}_{\rm P}xx^{\prime}b^{3}+3\tilde{\sigma}_{\rm P}x^{2}b^{2}b^{\prime}-\tilde{\sigma}_{\rm H}x^{2}b^{2}b_{\phi}-s\tilde{\sigma}_{\rm H}^{\prime}x^{2}b^{2}b_{\phi} \nonumber \\
& &-2s\tilde{\sigma}_{\rm H}xx^{\prime}b^{2}b_{\phi}-2s\tilde{\sigma}_{\rm H}x^{2}bb^{\prime}b_{\phi}-s\tilde{\sigma}_{\rm H}x^{2}b^{2}b_{\phi}^{\prime}\big]\bigg)\bigg/\left[\left(\tilde{\sigma}_{\rm O}-\tilde{\sigma}_{\rm P}\right)bb_{z}+\tilde{\sigma}_{\rm P}x^{2}b^{3}-s\tilde{\sigma}_{\rm H}x^{2}b^{2}b_{\phi}\right]^{2} \nonumber \,,
\end{eqnarray}

\bigskip

\noindent and

\begin{eqnarray}
B^{\prime} & = & \bigg(\left[e_{\phi}^{\prime}+v_{z}^{\prime}b_{r}+v_{z}b_{r}^{\prime}-v_{r}^{\prime}b_{z}-v_{r}b_{z}^{\prime}\right]\left[-\tilde{\sigma}_{\rm P}x^{2}bb_{r}b_{\phi}+s\tilde{\sigma}_{\rm H}x^{2}b^{2}b_{r}+\tilde{\sigma}_{\rm H}b_{r}^{2}\right] \nonumber \\
& & + e_{\rm{c}\it{\phi}}\left[-\tilde{\sigma}_{\rm P}^{\prime}x^{2}bb_{r}b_{\phi}-2\tilde{\sigma}_{\rm P}xx^{\prime}bb_{r}b_{\phi}-\tilde{\sigma}_{\rm P}x^{2}b^{\prime}b_{r}b_{\phi}-\tilde{\sigma}_{\rm P}x^{2}bb_{r}^{\prime}b_{\phi}-\tilde{\sigma}_{\rm P}x^{2}bb_{r}b_{\phi}^{\prime} \right. \nonumber \\
& & \left. +\tilde{\sigma}_{\rm H}x^{2}b^{2}b_{r}+s\tilde{\sigma}_{\rm H}^{\prime}x^{2}b^{2}b_{r}+2s\tilde{\sigma}_{\rm H}xx^{\prime}b^{2}b_{r}+2s\tilde{\sigma}_{\rm H}x^{2}bb^{\prime}b_{r}+s\tilde{\sigma}_{\rm H}x^{2}b^{2}b_{r}^{\prime} \right. \nonumber \\
& & \left. +\tilde{\sigma}_{\rm H}^{\prime}b_{r}^{2}+2\tilde{\sigma}_{\rm H}b_{r}b_{r}^{\prime}\right]-\left[\frac{1}{\zeta}-1\right]\big[x^{\prime}bb_{r}b_{\phi}b_{z}+xb^{\prime}b_{r}b_{\phi}b_{z}+xbb_{r}^{\prime}b_{\phi}b_{z} \nonumber\\
& & +xbb_{r}b_{\phi}^{\prime}b_{z}+xbb_{r}b_{\phi}b_{z}^{\prime}\big]\bigg)\bigg/\left[\left(\tilde{\sigma}_{\rm O}-\tilde{\sigma}_{\rm P}\right)bb_{z}+\tilde{\sigma}_{\rm P}x^{2}b^{3}-s\tilde{\sigma}_{\rm H}x^{2}b^{2}b_{\phi}\right] \nonumber \\
& & -\bigg(\left[e_{\rm{c}\it{\phi}}\left(-\tilde{\sigma}_{\rm P}x^{2}bb_{r}b_{\phi}+s\tilde{\sigma}_{\rm H}x^{2}b^{2}b_{r}+\tilde{\sigma}_{\rm H}b_{r}^{2}\right)-\left(\frac{1}{\zeta}-1\right)xbb_{r}b_{\phi}b_{z}\right] \big[\left(\tilde{\sigma}_{\rm O}^{\prime}-\tilde{\sigma}_{\rm P}^{\prime}\right)bb_{z} \nonumber \\
& & +\left(\tilde{\sigma}_{\rm O}-\tilde{\sigma}_{\rm P}\right)b^{\prime}b_{z}+\left(\tilde{\sigma}_{\rm O}-\tilde{\sigma}_{\rm P}\right)bb_{z}^{\prime}+\tilde{\sigma}_{\rm P}x^{2}b^{3}+2\tilde{\sigma}_{\rm P}xx^{\prime}b^{3}+3\tilde{\sigma}_{\rm P}x^{2}b^{2}b^{\prime}-\tilde{\sigma}_{\rm H}x^{2}b^{2}b_{\phi}-s\tilde{\sigma}_{\rm H}^{\prime}x^{2}b^{2}b_{\phi} \nonumber \\
& &-2s\tilde{\sigma}_{\rm H}xx^{\prime}b^{2}b_{\phi}-2s\tilde{\sigma}_{\rm H}x^{2}bb^{\prime}b_{\phi}-s\tilde{\sigma}_{\rm H}x^{2}b^{2}b_{\phi}^{\prime}\big]\bigg)\bigg/\left[\left(\tilde{\sigma}_{\rm O}-\tilde{\sigma}_{\rm P}\right)bb_{z}+\tilde{\sigma}_{\rm P}x^{2}b^{3}-s\tilde{\sigma}_{\rm H}x^{2}b^{2}b_{\phi}\right]^{2} \nonumber \,.
\end{eqnarray}

Substituting Equation (\ref{eq:ezerprime}) into Equation (\ref{eq:induct2a}), we have

\begin{displaymath}
e_{r}^{\prime}+s\left(Ce_{r}^{\prime}+D\right) = \left(\frac{1}{\xi}-\frac{5}{2}\right)\left(\frac{x^{\prime}}{x}\left[e_{r}+e_{z}s\right]+e_{z}\right) \nonumber \,,
\end{displaymath}

\noindent which gives

\begin{equation}
e_{r}^{\prime} = \frac{\left(\frac{1}{\xi}-\frac{5}{2}\right)\left(\frac{x^{\prime}}{x}\left[e_{r}+e_{z}s\right]+e_{z}\right)-Ds}{1+Cs} \,.
\label{eq:erprime}
\end{equation}

\newpage

\section{Taylor Series Expansions from the Midplane}\label{apptaylor}

As described in Section~\ref{midgen}, the midplane is a singular point of the disk equations.  The procedure adopted here for dealing with this issue is to calculate Taylor series expansions of the variables of integration from the midplane to a small distance $\Delta s = 0.001$ above the midplane before beginning numerical integration of the disk equations.  The Taylor series expansions for a general, non-ideal disk described using the conductivity tensor formalism are presented in Section~\ref{taylorgen}.  The case of a disk in the ambipolar diffusion regime with constant neutral--field coupling parameter $\Upsilon$ is covered in Section~\ref{taylorAD}.

\subsection{General Case}\label{taylorgen}

Differentiating Equation (\ref{eq:xprime}) once with respect to $s$, we have $x^{\prime\prime} = b_{r}^{\prime}x^{3}+3b_{r}x^{2}x^{\prime} = b_{r}^{\prime}x^{3}+3b_{r}^{2}x^{2}$; dropping terms that vanish at the midplane gives
\begin{equation}
x_{0}^{\prime\prime} = b_{r,0}^{\prime} \,.
\label{eq:ppx0}
\end{equation}

Differentiating Equation (\ref{eq:rhoprime}) once with respect to $s$ and dropping terms that vanish at the midplane gives
\begin{eqnarray}
\tilde{\rho}_{0}^{\prime\prime} & = & \left[-\frac{1}{\theta}\right]\bigg[\left(2-\frac{2}{\zeta}\right)v_{r,0}\left(v_{z,0}^{\prime}-v_{r,0}\right)+\left(4-\frac{2}{\zeta}\right)\theta\left(1+x_{0}^{\prime\prime}\right) \nonumber \\
& & +v_{\phi,0}^{2}+\nu\left(j_{\phi,0}\left[b_{r,0}^{\prime}+1\right]-b_{\phi,0}^{\prime}j_{r,0}\right)\bigg] \,.
\label{eq:pprho0}
\end{eqnarray}

Differentiating Equation (\ref{eq:induct2a}) once with respect to $s$ and dropping terms that vanish at the midplane gives
\begin{equation}
e_{r,0}^{\prime\prime} + e_{z,0}^{\prime} = \left(\frac{1}{\xi}-\frac{5}{2}\right)\left(x_{0}^{\prime\prime}e_{r,0}+e_{z,0}^{\prime}\right) \,.
\label{eq:pper0}
\end{equation}

\noindent Differentiating the $z$ component of Equation (\ref{eq:ec2}) gives $e_{z,0}^{\prime} = e_{{\rm c}z,0}^{\prime} + v_{\phi,0}^{\prime}b_{r,0}^{\prime}-v_{r,0}b_{\phi,0}^{\prime} = e_{{\rm c}z\parallel,0}^{\prime} + e_{{\rm c}z\perp,0}^{\prime} + v_{\phi,0}^{\prime}b_{r,0}^{\prime}-v_{r,0}b_{\phi,0}^{\prime}$.  Differentiating Equation (\ref{eq:ecperp}) yields $e_{{\rm c}z\perp,0}^{\prime} = -b_{r,0}^{\prime}e_{{\rm c}r,0}-b_{\phi,0}^{\prime}e_{{\rm c}\phi,0}$.  Differentiating Equation (\ref{eq:ampz2}) and Equation (\ref{eq:ohm2z}) gives $j_{z,0}^{\prime} = \left(1/\zeta-2\right)b_{\phi,0}^{\prime}$ and $e_{{\rm c}z\parallel,0}^{\prime} = \left(j_{z,0}^{\prime}+\tilde{\sigma}_{\rm H,0}\left[b_{\phi,0}^{\prime}e_{{\rm c}r,0}-b_{r,0}^{\prime}e_{{\rm c}\phi,0}\right]-\tilde{\sigma}_{\rm P,0}e_{{\rm c}z\perp,0}^{\prime}\right)/\tilde{\sigma}_{\rm O,0}$, respectively.  These relations can be used to solve for $e_{r,0}^{\prime\prime}$ in Equation (\ref{eq:pper0}).

Calculating $v_{r,0}^{\prime\prime}$ and $v_{\phi,0}^{\prime\prime}$ is somewhat more complex.  Differentiating Equations (\ref{eq:mom2r}) and (\ref{eq:mom2phi}) twice with respect to $s$ and dropping terms that vanish at the midplane gives
\begin{eqnarray}
2\left(v_{z,0}^{\prime}-v_{r,0}\right)v_{r,0}^{\prime\prime}-2\theta\tilde{\rho}_{0}^{\prime\prime} & = & v_{r,0}v_{r,0}^{\prime\prime} + 2v_{\phi,0}v_{\phi,0}^{\prime\prime}-v_{r,0}\left(v_{z,0}^{\prime}-v_{r,0}\right)x_{0}^{\prime\prime}-3 \nonumber \\
& & +2\left(4-\frac{2}{\zeta}\right)\theta x_{0}^{\prime\prime}+\nu x_{0}^{\prime\prime}j_{\phi,0}-\nu \tilde{\rho}_{0}^{\prime\prime}j_{\phi,0}+\nu j_{\phi,0}^{\prime\prime}-2\nu j_{z,0}^{\prime}b_{\phi}^{\prime}
\label{eq:ppvr0}
\end{eqnarray}

\noindent and

\begin{eqnarray}
2\left(v_{z,0}^{\prime}-v_{r,0}\right)v_{\phi,0}^{\prime\prime} & = & -\frac{v_{r,0}^{\prime\prime} v_{\phi,0}}{2} - \frac{v_{r,0} v_{\phi,0}^{\prime\prime}}{2} + 2v_{\phi,0}v_{\phi,0}^{\prime\prime}-v_{\phi,0}\left(v_{z,0}^{\prime}-v_{r,0}\right)x_{0}^{\prime\prime} \nonumber \\
& & -\nu x_{0}^{\prime\prime}j_{r,0}+\nu \tilde{\rho}_{0}^{\prime\prime}j_{r,0}-\nu j_{r,0}^{\prime\prime}+2\nu j_{z,0}^{\prime}b_{r}^{\prime} \,.
\label{eq:ppvphi0}
\end{eqnarray}

\noindent Expressions for the unknown quantities $j_{r,0}^{\prime\prime}$ and $j_{\phi,0}^{\prime\prime}$ can be derived by differentiating Equations (\ref{eq:ampr2}) and (\ref{eq:ampphi2}) twice with respect to $s$ and dropping terms that vanish at the midplane, which gives

\begin{eqnarray}
j_{r,0}^{\prime\prime} & = & \tilde{\sigma}_{\rm O,0}e_{{\rm c}r\parallel,0}^{\prime\prime} - \tilde{\sigma}_{\rm H,0}^{\prime\prime} e_{\rm c\phi,0} + \tilde{\sigma}_{\rm H,0}b_{0}^{\prime\prime}e_{\rm c\phi,0} + \tilde{\sigma}_{\rm H,0}\left(2b_{\phi,0}^{\prime}e_{{\rm c}z\perp,0}^{\prime}-e_{\rm c\phi\perp,0}^{\prime\prime}\right)\nonumber \\
& & + \tilde{\sigma}_{\rm P,0}^{\prime\prime} e_{{\rm c}r,0}+ \tilde{\sigma}_{\rm P,0} e_{{\rm c}r\perp,0}^{\prime\prime}
\label{eq:ppjr0}
\end{eqnarray}

\noindent and 

\begin{eqnarray}
j_{\phi,0}^{\prime\prime} & = & \tilde{\sigma}_{\rm O,0}e_{\rm c\phi\parallel,0}^{\prime\prime} + \tilde{\sigma}_{\rm H,0}^{\prime\prime} e_{{\rm c}r,0} - \tilde{\sigma}_{\rm H,0}b_{0}^{\prime\prime}e_{{\rm c}r,0} + \tilde{\sigma}_{\rm H,0}\left(e_{{\rm c}r\perp,0}^{\prime\prime}-2b_{r,0}^{\prime}e_{{\rm c}z\perp,0}^{\prime}\right)\nonumber \\
& & + \tilde{\sigma}_{\rm P,0}^{\prime\prime} e_{\rm c\phi,0}+ \tilde{\sigma}_{\rm P,0} e_{\rm c\phi\perp,0}^{\prime\prime} \,.
\label{eq:ppjphi0}
\end{eqnarray}

\noindent We have $b_{0}^{\prime\prime} = b_{r,0}^{\prime2} + b_{\phi,0}^{\prime2}$, since $b_{z,0}^{\prime} = b_{z,0}^{\prime\prime} = 0$.  Differentiating Equation (\ref{eq:ecpar}) twice gives the quantities $e_{{\rm c}r\parallel,0}^{\prime\prime}$ and $e_{\rm c\phi\parallel,0}^{\prime\prime}$ as $e_{{\rm c}r\parallel,0}^{\prime\prime} = 2b_{r,0}^{\prime}e_{{\rm c}z\parallel,0}^{\prime}$ and
$e_{\rm c\phi\parallel,0}^{\prime\prime} = 2b_{\phi,0}^{\prime}e_{{\rm c}z\parallel,0}^{\prime}$.  The quantities $e_{{\rm c}r\perp,0}^{\prime\prime}$ and $e_{\rm c\phi\perp,0}^{\prime\prime}$ are obtained by differentiating Equation (\ref{eq:ecparperp}) twice to obtain $e_{{\rm c}r\perp,0}^{\prime\prime} = e_{{\rm c}r,0}^{\prime\prime} - e_{{\rm c}r\parallel,0}^{\prime\prime}$ and $e_{\rm c\phi\perp,0}^{\prime\prime} = e_{\rm c\phi,0}^{\prime\prime} - e_{\rm c\phi\parallel,0}^{\prime\prime}$.  Finally, differentiating Equation (\ref{eq:ec2}) twice gives $e_{{\rm c}r,0}^{\prime\prime} = e_{r,0}^{\prime\prime} + v_{\phi,0}^{\prime\prime} - 2v_{z,0}^{\prime}b_{\phi,0}^{\prime}$ and $e_{{\rm c}\phi,0}^{\prime\prime} = 2v_{z,0}^{\prime}b_{\phi,0}^{\prime} - v_{r,0}^{\prime\prime}$.  Plugging these relations back into Equation (\ref{eq:ppvr0}) and Equation (\ref{eq:ppvphi0}) gives two equations that can be solved for $v_{r,0}^{\prime\prime}$ and $v_{\phi,0}^{\prime\prime}$ in terms of known quantities.

\subsection{Ambipolar Diffusion Case with Constant $\Upsilon$}\label{taylorAD}
As in the general conductivity case, $x_{0}^{\prime\prime} = b_{r,0}^{\prime}$.

Differentiating the relation \boldmath$B$\unboldmath$ \cdot \left(\mbox{\boldmath$V$}_{\rm i} - \mbox{\boldmath$V$}\right) = 0$ once and evaluating at the midplane gives $v_{\rm{i}\mathit{z},0}^{\prime} = v_{z,0}^{\prime} + b_{r,0}^{\prime}\left(
v_{\rm{i}\mathit{r},0} - v_{r,0}\right) + b_{\phi,0}^{\prime} \left(v_{\rm{i}\mathit{\phi},0} - v_{\phi,0}\right)$.  Differentiating Equation (\ref{eq:virz}) twice and evaluating at the midplane yields

\begin{equation}
v_{\rm{i}\mathit{r},0}^{\prime\prime} = 2b_{r,0}^{\prime}v_{\rm{i}\mathit{z},0}^{\prime} - \left(\frac{1}{\zeta}-\frac{5}{2}\right)\epsilon_{\rm B}\theta^{\frac{1}{2}}b_{r,0}^{\prime} \,.
\label{eq:ppvir0ad}
\end{equation}

Differentiating the $z$ component of Equation (\ref{eq:moment}) (cf. Equation (12) of L96) once and evaluating at the midplane gives
\begin{equation}
\tilde{\rho}_{0}^{\prime\prime} = \left(v_{z,0}^{\prime}\left[\frac{3v_{r,0}}{2}-v_{z,0}^{\prime}\right]-\left[4-\frac{2}{\zeta}\right]\theta x_{0}^{\prime\prime}-1+\Upsilon_{0}\left[v_{\rm{i}\mathit{z},0}^{\prime}-v_{z,0}^{\prime}\right]\right)/\theta \,.
\label{eq:pprho0ad}
\end{equation}

Differentiating Equation (\ref{eq:virphiprime}) three times and evaluating at the midplane yields
\begin{equation}
v_{\rm{i}\mathit{\phi},0}^{\prime\prime} =  \frac{b_{\phi,0}^{\prime}}{b_{r,0}^{\prime}}v_{\rm{i}\mathit{r},0}^{\prime\prime} + \left(v_{\rm{i}\mathit{\phi},0}b_{r,0}^{\prime}-v_{\rm{i}\mathit{r},0}b_{\phi,0}^{\prime}\right) \,.
\label{eq:ppviphi0ad}
\end{equation}

Finally, differentiating the $r$ and $\phi$ components of Equation (\ref{eq:moment}) (cf. Equations (13) and (14) of L96) twice and evaluating at the midplane gives

\begin{eqnarray}
\left(3v_{r,0}-2v_{z,0}^{\prime}-\Upsilon_{0}\right)v_{r,0}^{\prime\prime}+2v_{\phi,0}v_{\phi,0}^{\prime\prime} & = & v_{r,0}\left(v_{z,0}^\prime-v_{r,0}\right)x_{0}^{\prime\prime}-2\theta\tilde{\rho}_{0}^{\prime\prime}-2\left(4-\frac{2}{\zeta}\right)\theta x_{0}^{\prime\prime} \nonumber \\
& & -x_{0}^{\prime\prime}+3-\Upsilon_{0}v_{\rm{i}\mathit{r},0}^{\prime\prime}+\Upsilon_{0}^{\prime\prime}\left(v_{r,0}^{\prime\prime}-v_{\rm{i}\mathit{r},0}^{\prime\prime}\right)
\label{eq:ppvr0ad}
\end{eqnarray}

\noindent and

\begin{equation}
\frac{v_{\phi,0}}{2}v_{r,0}^{\prime\prime}+\left(2v_{z,0}^{\prime\prime}-\frac{3}{2}v_{r,0}+\Upsilon_{0}\right)v_{\phi,0}^{\prime\prime} = \Upsilon_{0}v_{\rm{i}\mathit{\phi},0}^{\prime\prime} + \Upsilon_{0}^{\prime\prime}\left(v_{\rm{i}\mathit{\phi},0} - v_{\phi,0}\right)-v_{\phi,0}\left(v_{z,0}^{\prime}-v_{r,0}\right)x_{0}^{\prime\prime} \,.
\label{eq:ppvphi0ad}
\end{equation}

\noindent These equations can be solved for $v_{r,0}^{\prime\prime}$ and $v_{\phi,0}^{\prime\prime}$ in terms of known quantities.

\end{document}